\documentclass{sig-alternate}

\usepackage[hyphens]{url}
\usepackage{graphicx}
\usepackage{algorithm,algpseudocode,subfigure,amsmath}
\begin{document}
\conferenceinfo{}{}
\CopyrightYear{} % Allows default copyright year (20XX) to be over-ridden - IF NEED BE.
%\crdata{0-12345-67-8/90/01}  % Allows default copyright data (0-89791-88-6/97/05) to be over-ridden - IF NEED BE.

\title{Hiding in Plain Sight: The Anatomy of Malicious\\ Facebook {\ttlit Pages}}

\numberofauthors{1}
\author{
\alignauthor
Prateek Dewan, Ponnurangam Kumaraguru\\
       \affaddr{Cybersecurity Education and Research Centre (CERC@IIITD)}\\
       \affaddr{Indraprastha Institute of Information Technology - Delhi (IIITD)}\\
       \affaddr{New Delhi, India}\\
       \email{\{prateek, pk\}@iiitd.ac.in}
}
\date{17 October 2015}

\maketitle
\begin{abstract}

Facebook is the world's largest Online Social Network, having more than 1 billion users. Like most other social networks, Facebook is home to various categories of hostile entities who abuse the platform by posting malicious content. In this paper, we identify and characterize Facebook pages that engage in spreading URLs pointing to malicious domains. We used the Web of Trust API to determine domain reputations of URLs published by pages, and identified 627 pages publishing untrustworthy information, misleading content, adult and child unsafe content, scams, etc. which are deemed as ``Page Spam'' by Facebook, and do not comply with Facebook's community standards.

Our findings revealed dominant presence of politically polarized entities engaging in spreading content from untrustworthy web domains. Anger and religion were the most prominent topics in the textual content published by these pages. We found that at least 8\% of all malicious pages were dedicated to promote a single malicious domain. Studying the temporal posting activity of pages revealed that malicious pages were more active than benign pages. We further identified collusive behavior within a set of malicious pages spreading adult and pornographic content. We believe our findings will enable technologists to devise efficient automated solutions to identify and curb the spread of malicious content through such pages. To the best of our knowledge, this is the first attempt in literature, focused exclusively on characterizing malicious Facebook pages.

\end{abstract}
%\category{H.4}{Information Systems Applications}{Miscellaneous}

%\terms{Theory}


\section{Introduction}

Online Social Networks (OSNs) are an integral part of the World Wide Web today. Internet users around the world use OSNs as primary sources to consume news, updates, and information about events around the world. Twitter is the most preferred platform for breaking news, while Facebook leads the way in terms of the number of users who consume news from the social network~\cite{Michael-Barthel:2015}. However, given the enormous volume, it is hard to moderate all content that is generated and shared on OSNs. This enables hostile entities to generate and promote all sorts of malicious content and pollute the information stream for monetary gains, or to simply compromise system reputation. Such activity not only degrades user experience, but also violates the terms of service of most OSN platforms.

\begin{figure}[!b]
\begin{center}
\fbox{\includegraphics[scale=0.37]{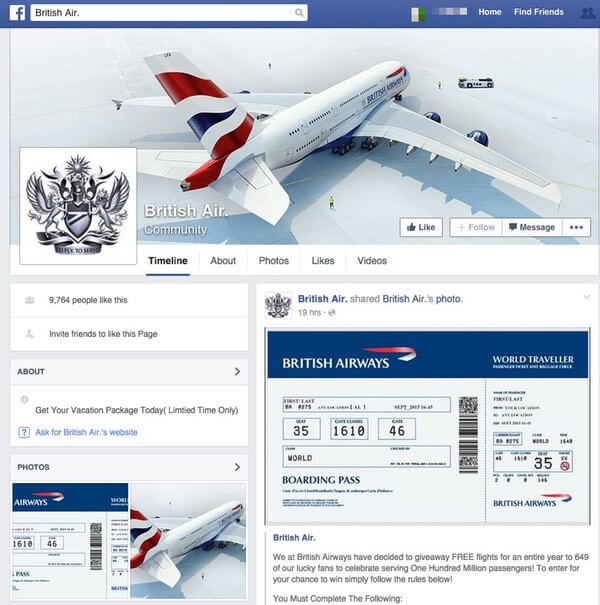}}
\end{center}
\caption{Fake British Air Facebook page offering free flights for a year in return for liking, commenting on, and sharing their post.}
\label{fig:fbpage_screenshot}
\end{figure}

Multiple researchers in the past have proposed automated techniques to identify hostile user accounts on OSNs that engage in spreading malware, unsolicited spam, and phishing~\cite{benevenuto2010detecting,lee2010uncovering,stringhini2010detecting,thomas2011suspended}, or collude to artificially inflate the social capital of other entities~\cite{stringhini2012poultry,viswanath2014towards}. However, in a recent study, we discovered the presence of malicious Facebook \emph{pages}, which have not been studied in detail in the past%. We found greater participation of Facebook \emph{pages} in posting malicious URLs (21\%) as compared to posting benign URLs (10\%)
~\cite{dewan2015towards}. Certain security experts and news sources have also acknowledged the presence of malicious \emph{pages} on Facebook, set up intentionally to spread phishing, scams, and other types of malicious content. For example, a group of scammers set up a fake British Airways page, offering free flights to customers for a year (Figure~\ref{fig:fbpage_screenshot}). The page asked users to \emph{share} a photo, \emph{like} the page and leave a \emph{comment} to claim their free flights.~\footnote{\url{https://grahamcluley.com/2015/09/british-airways-isnt-giving-away-free-flights-year-facebook-scam/}} In another similar incident, an international football player's name was used as bait to set up a Facebook page, and users were asked to sign a fake petition.~\footnote{\url{http://www.marca.com/2014/07/18/en/football/barcelona/1405709402.html}} Scammers can use such events to gather large audience for their pages, and then bombard users with other unwanted, spammy promotions, and potentially dodgy links that could lead to a malware infection or users being phished. It has been claimed that Facebook pages spam is a \$200 million business.~\footnote{\url{http://mashable.com/2013/08/29/facebook-fan-pages-spam-200-million-business/}}

Facebook pages are an important and integral part of the Facebook ecosystem, that offer a free platform for promotion of businesses, brands and organizations.~\footnote{\url{https://www.facebook.com/help/174987089221178}} From an attacker's perspective, Facebook pages are potentially lucrative tools to gather large audiences and target all of them at once. %Some recent events and research have established the participation of Facebook pages as tools for spreading malicious content on Facebook~\cite{Marca.com:2014,Cluley:2015,dewan2015towards}. 
However, past research on malicious entities on Facebook is largely restricted to identifying user profiles~\cite{ahmed2012mcl,stringhini2010detecting}. In this paper, we focus strictly on malicious Facebook pages. We identify and characterize a set of 627 pages that published one or more malicious URLs in their most recent 100 posts. We go beyond traditional classes of malicious content like phishing and spam, and also study pages which spread untrustworthy information, hate speech, nudity, misleading claims. Such content is also deemed as malicious by community standards~\cite{Facebook.com:2015} and ``Page Spam'' definitions established by Facebook. We identify malicious pages by looking up the Web of Trust (WOT) domain reputations of URLs published by pages, and filter out pages posting URLs from low-reputation domains (in terms of trustworthiness, child safety, etc.). Then, we explore who are the entities behind these pages, and characterize and compare the behavior of malicious pages with benign pages. Our broad findings are as follows:

\begin{itemize}
\item {\bf Politically polarized malicious entities}: We identified numerous politically polarized entities which dominated our dataset of malicious pages, and published URLs from untrustworthy web domains.

\item {\bf Dedicated pages for malicious domains}: We found that at least 8\% of pages in our dataset were dedicated to promoting content from only one malicious web domain. This behavior indicated that some malicious entities intentionally set up Facebook pages dedicated to spreading their own content.

\item {\bf Malicious pages were more active}: We found that malicious pages were more active (in terms of posting) than benign pages; the number of malicious pages that were active daily was over three times the number of benign pages that were active daily.

\item {\bf Malicious pages showed collusive behavior}: We found presence of collusive behavior within malicious pages in our dataset; malicious pages engaged in promoting (\emph{liking, commenting on}, and \emph{sharing}) each others' content.

%\item We found numerous similarities between the behavior of malicious and benign pages, which makes it hard to differentiate malicious pages from benign ones using automated means. These similarities enable malicious pages to hide in plain sight and continue to operate and thrive.

\end{itemize}

To the best of our knowledge, this is the first attempt in literature, focused exclusively on characterizing malicious Facebook pages. %We believe that this work will enable researchers to better understand the landscape of malicious pages on Facebook. 
The rest of the paper is structured as follows. Section~\ref{sec:background} gives the background and scope of our research, and explains the data collection process. Characterization and analysis of malicious pages make up Section~\ref{sec:analysis}. Section~\ref{sec:discussion} contains the discussion and implications of our results. Related work is discussed in Section~\ref{sec:relatedwork}. We finally conclude our work in Section~\ref{sec:conclusion}.

%- we wanted to see if pages accidentally spread malicious links or intentionally (dedicated to one BAD domain). <-- Motivation for 4.1

\section{Background} \label{sec:background}

%Most Online Social Networks (OSNs) have two prominent types of entities using the network. These include regular users with comparable in-degree and out-degree connections with other users, and celebrity users with a much larger in-degree as compared to the out-degree. Celebrity users tend to be followed by masses and have a much wider audience as compared to regular users. 
Some OSNs like Twitter and Instagram do not have restrictions on the number of connections an entity (user) can have. %, and treat regular and celebrity users the same. 
However, other networks like Facebook pose a restriction on the number of connections a user can have, and provide \emph{pages} to enable large following for celebrities, groups, businesses, etc. A Facebook \emph{page} can have multiple administrators managing the \emph{page} under the same name, without the audience knowing. This allows \emph{pages} to have a higher degree of interaction with its audience and keeping it more active as compared to a single user profile. Also, the most popular \emph{page} on Facebook has an audience of 514,654,759 fans (\emph{likes}), which is approximately 100,000 times larger than the maximum audience (\emph{friends}) a Facebook user profile can have (5,000 friends).~\footnote{As on September 22, 2015. \url{http://fanpagelist.com/category/top_users/}}~\footnote{Although user profiles on Facebook have a \emph{follow} option, \emph{followers} of a user can only view posts which are public.} %Moreover, all content generated by a Facebook page is public, and can be viewed by anyone on the Internet. 
Our past research has shown greater participation of Facebook pages in posting malicious URLs as compared to posting benign URLs~\cite{dewan2015towards}. Such inflated malicious activity and wide reach of Facebook pages make them a vital aspect to study independently in detail. 

\subsection{Scope}

The definition and scope of what should be labeled as ``malicious content'' on the Internet has been constantly evolving since the birth of the Internet. Researchers have been studying malicious content in the form of spam and phishing for over two decades. With respect to Online Social Networks, state-of-the-art techniques have become efficient in automatically detecting spam campaigns~\cite{gao2010detecting,zhang2012detecting}, and phishing~\cite{aggarwal2012phishari} without human involvement. However, new classes of malicious content pertaining to appropriateness, authenticity, trustworthiness, and credibility of content have emerged in the recent past. Some researchers have studied these classes of malicious content on OSNs and shown their implications in the real world~\cite{castillo2011information,gupta2012credibility,gupta2012evaluating,mendoza2010twitter}. All of these studies, however, resorted to human expertise to identify untrustworthy and inappropriate content and establish ground truth, due to the absence of efficient automated techniques to identify such content. We aim to study a similar class of malicious content pertaining to trustworthiness and appropriateness in this work, which currently requires human expertise to identify.
%There exists a wide variety of malicious activity that has been studied on OSNs in the past. These include studies focusing on spam campaigns~\cite{gao2010detecting}, content credibility~\cite{castillo2011information}, social malware~\cite{rahman2012efficient}, URL spam~\cite{grier2010spam}, fake likes / followers~\cite{stringhini2012poultry,viswanath2014towards}, and many more. We do not intend to study all such behavior. In this study, we focus our analysis on studying Facebook pages who post malicious URLs. 
In particular, we look at Facebook pages that generate content deemed as malicious by Facebook's community standards and definitions of ``Page Spam''. Facebook defines ``Page Spam'' as pages that \emph{confuse, mislead, surprise or defraud people}.~\footnote{\url{https://www.facebook.com/help/116053525145846}} Additionally, pages that are misleading, deceptive, or otherwise misrepresent the prize or any other aspect of promotion are considered as ``Page Spam''. Facebook has also established community standards to protect users against nudity, hate speech, violence and graphic content, fraud, sexual violence etc.~\cite{Facebook.com:2015}.

Most URL blacklists (Google Safebrowsing, PhishTank, SURBL, SpamHaus, etc.) used in past research are useful to identify more obvious threats like phishing, spam, malware, etc. These blacklists are reasonably efficient, but are not capable of identifying websites engaging in spreading misleading information, hate speech, unethical claims, nudity, etc. which are considered as Facebook's community standards and ``Page Spam'' definitions. In order to obtain ground truth for such kind of malicious content, we used the Web of Trust (WOT) API. WOT is a website reputation and review service that helps people make informed decisions about whether to trust a website or not~\cite{WOT:2014}. WOT is based on a unique crowdsourcing approach that collects ratings and reviews from a global community of millions of users who rate and comment on websites based on their personal experiences. The community-powered approach enables WOT to protect users against threats that only the human eye can spot such as scams, unreliable web stores, misleading websites, nudity, and questionable content, which largely overlaps with Facebook definitions of spam.

We understand that WOT ratings are obtained through crowd sourcing, and may seem to suffer from biases. However, WOT states that in order to keep ratings more reliable, the system tracks each user's rating behavior before deciding how much it trusts the user. In addition, the meritocratic nature of WOT makes it far more difficult for spammers to abuse. We now discuss our data collection methodology in detail.

\subsection{Dataset} \label{sec:dataset}

We collected a dataset of 4.4 million public posts from Facebook between April 2013 and July 2014, and identified 11,217 posts containing malicious URLs by looking up six URL blacklists~\cite{dewan2015towards}. Table~\ref{tab:descstats} shows the descriptive statistics of this dataset. We used the same dataset of Facebook posts to obtain a true positive dataset of pages posting malicious URLs. We rescanned the 1,696 pages posting malicious URLs (as identified in our previous work~\cite{dewan2015towards}) to collect their page information through the Facebook Graph API in August 2015, and found that 418 out of the 1,696 pages did not exist. For the remaining pages, we collected all posts published by the pages using the \emph{/page-id/posts} endpoint of the Graph API.~\footnote{\url{https://developers.facebook.com/docs/graph-api/reference/page/feed}} We also collected all \emph{likes}, \emph{comments}, and \emph{shares} on the most recent 100 posts published by these pages. We then looked up the WOT API for all URL domains present in the most recent 100 posts, and found that 627 pages published one or more malicious URLs. We considered these pages as malicious. Interestingly, we found 31 malicious pages in our dataset that were verified by Facebook.~\footnote{\url{https://www.facebook.com/help/196050490547892}}

Due to time constraints, we had to restrict the WOT API lookups and collection of \emph{likes}, \emph{comments}, and \emph{shares} to the most recent 100 posts only. Researchers have adopted similar methodology in the past, where they identified malicious URLs in the 40 most recent Tweets of users to obtain a ground truth dataset of spammers on Twitter~\cite{yang2011free}.
 
\begin{table}[!h]
\begin{center}
    \begin{tabular}{l|r}
    \hline
    Total posts				& 4,465,371	\\ \hline
    Total unique users			& 2,983,707	\\ \hline
    Total unique pages			& 390,246	\\ \hline
    Unique posts with URLs		& 1,222,137	\\ \hline
    Total unique URLs			& 480,407	\\ \hline
    Unique posts with malicious URLs	& 11,217	\\ \hline
    Unique users posting malicious URLs	& 6,286		\\ \hline
    Unique pages posting malicious URLs	& 1,696		\\ \hline
    Unique malicious URLs		& 4,622		\\ \hline
    \end{tabular}
\caption{Descriptive statistics of complete dataset of Facebook posts collected by Dewan et al. over April 2013 - July 2014.}
\label{tab:descstats}
\end{center}
\end{table}

The WOT API returns a reputation score for a given domain. Reputations are measured for domains in several \emph{components}, for example, trustworthiness. For each {\fontfamily{qcr}\selectfont \{domain, component\}} pair, the system computes two values: a \emph{reputation} estimate and the \emph{confidence} in the reputation. Together, these indicate the amount of trust in the domain in the given component. A \emph{reputation} estimate of below 60 indicates \emph{unsatisfactory}. The WOT browser add-on requires a confidence value of $\geq$ 10 before it presents a warning about a website. We tested the domain of each URL in our dataset for \emph{Trustworthiness} and \emph{Child Safety} components. For our analysis, a URL was marked as malicious if both the aforementioned conditions were satisfied (Algorithm~1). In addition to reputations, the WOT rating system also computes categories for websites based on votes from users and third parties. We marked a URL as malicious if it fell under the \emph{Negative} (including malware, scams etc.) or \emph{Questionable} (including hate, incidental nudity etc.) category group.~\footnote{Exact category labels and description corresponding to \emph{Negative} and \emph{Questionable} categories can be found at \url{https://www.mywot.com/wiki/API}}

\begin{algorithm}
\small
\begin{algorithmic}
    \ForAll{pages}
        \ForAll{100 posts}
	    \ForAll{URL domains}
	    \State components $=$ GetComponentsFromWOT\_API
	        \ForAll{components}
		    \If {reputation $<$ 60 and confidence $\geq$ 10}
		        \State page $=$ malicious
			\State continue
		    \EndIf
	        \EndFor
	    \EndFor
        \EndFor
    \EndFor
\label{algo1}
\caption{Detecting pages posting malicious URLs using WOT reputation scores.}
\end{algorithmic}
\end{algorithm}

We also drew an equal random sample of 1,696 pages from the remaining benign pages in the dataset. Similar to our approach for identifying malicious pages, we collected all posts published by the random sample of benign pages. We rescanned the most recent 100 posts (including their \emph{likes}, \emph{comments}, and \emph{shares}) published by these pages and obtained 1,278 benign pages posting no malicious URLs in their most recent 100 posts. Table~\ref{tab:pagesData} shows the descriptive statistics of all Facebook pages in our dataset.

\begin{table}[!h]
\begin{center}
	\begin{tabular}{l|l|l}
	\hline
	Category						& Malicious	& Benign	\\ \hline
	No. of pages						& 627 (31)	& 1,278	(49)	\\
	All posts						& 9,341,420	& 3,890,101	\\
	Recent 100 posts					& 60,306	& 120,184	\\
	Recent 100 posts with URLs				& 55,233	& 92,980	\\
	Likes (recent 100 posts)\protect\footnotemark		& 3,447,669	& 31,680,263	\\
	Comments (recent 100 posts)				& 354,502	& 1,245,959	\\
	Shares (recent 100 posts)				& 507,964	& 1,012,151	\\ \hline
	\end{tabular}
\caption{Descriptive statistics of our dataset of Facebook pages. Numbers in the parentheses indicate verified pages.}
\label{tab:pagesData}
\end{center}
\end{table}

\footnotetext{Due to API rate limitations, we had to restrict our data collection to 50,000 \emph{likes} per post. We had 2 malicious and 291 benign posts which exceeded this limit.}

\begin{figure}[!h]
     \begin{center}
                \subfigure[Total number of posts published.]{%
                \label{fig:posts}
                \includegraphics[scale=0.19]{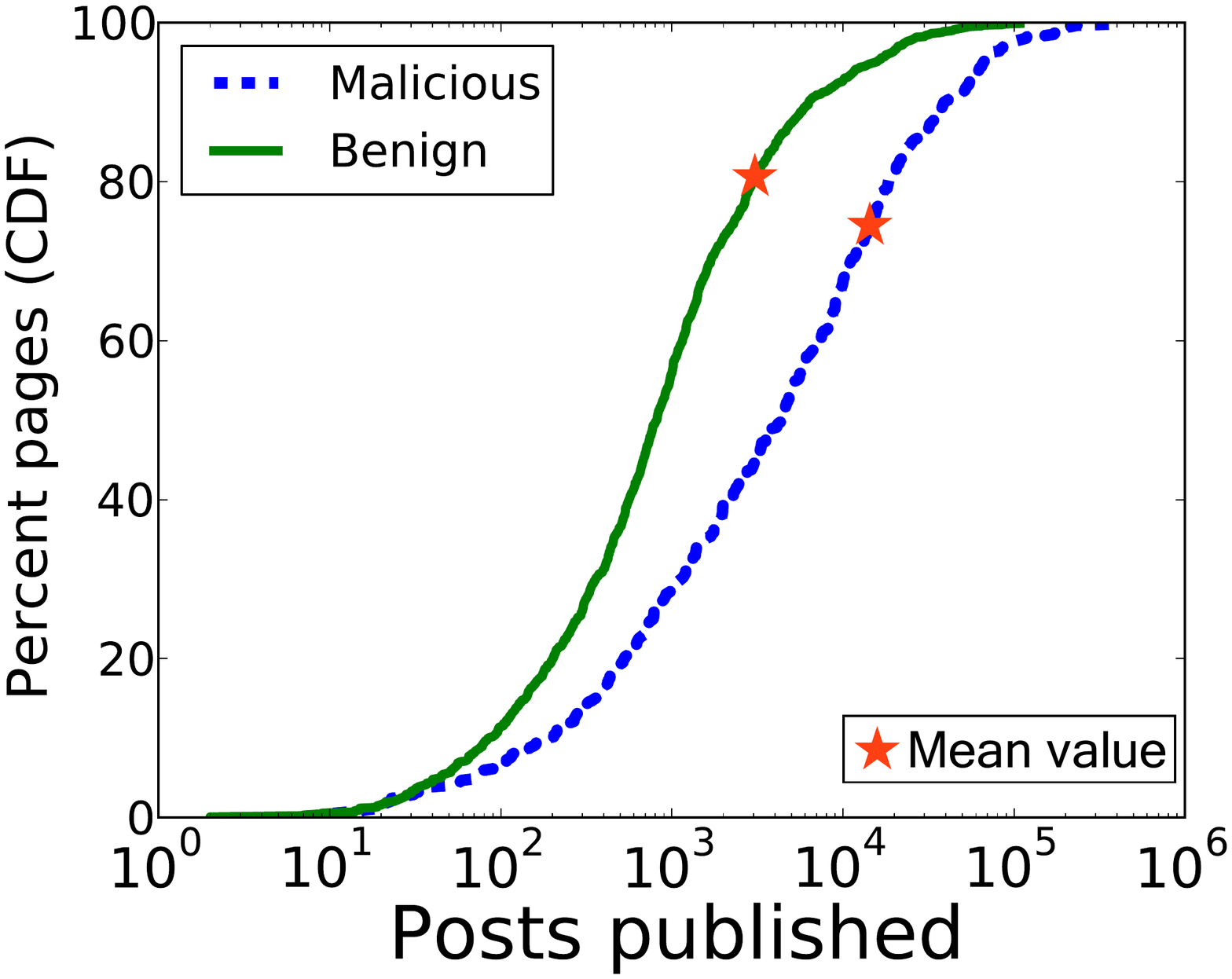}
        }
                \subfigure[Total number of page \emph{likes} gathered.]{
                \label{fig:likes}
                \includegraphics[scale=0.19]{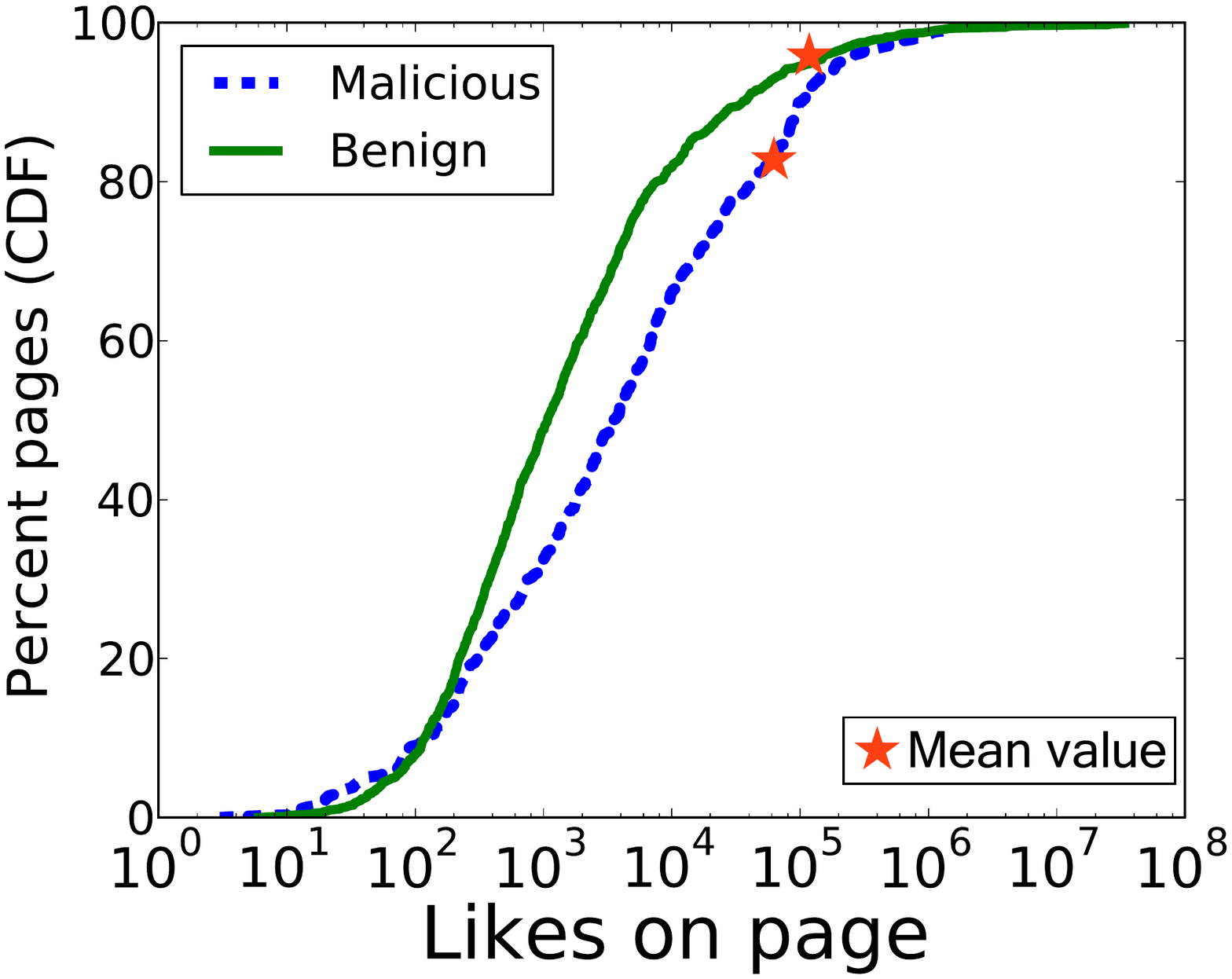}
        }
    \end{center}
\caption{Distribution of the total number of posts published, and number of page \emph{likes} gathered by pages in our dataset. Malicious pages published more posts than benign pages.}
   \label{fig:likesandposts}
\end{figure}

Figure~\ref{fig:likesandposts} shows the distribution of the number of posts published, and number of page \emph{likes} gathered by malicious and benign pages in our dataset. We observed that malicious pages published significantly more posts than benign pages ($\mu_{malicious}$ = 15,091.14 posts per page, $\mu_{benign}$ = 3,072.74 posts per page; Mann Whitney U statistic = 230474.0, \emph{p-value}<0.01). However, benign pages gathered more \emph{likes} than malicious pages ($\mu_{malicious}$ = 64,330.59 \emph{likes} per page, $\mu_{benign}$ = 112,250.36 \emph{likes} per page; Mann Whitney U statistic = 314109.5, \emph{p-value}<0.01). 

For the rest of the paper, we use the most recent 100 posts for all our analysis.

Table~\ref{tab:pagesPostsCatWise} provides a detailed description of the number of posts and pages along with their WOT categories in our malicious pages dataset of 627 pages. We found a total of 20,999 posts which contained one or more malicious URLs. These posts engaged a total of 675,162 unique users who \emph{liked}, \emph{commented}, or {shared} these posts. Interestingly, we found that spam and phishing (two of the most common types of malicious content studied in literature) were the least common types of malicious content in our dataset. Child unsafe content was the most common, followed by untrustworthy content.

\begin{table}[!ht]
\begin{center}
	\begin{tabular}{l|l|l}
	\hline
	WOT Response	& No. of pages	& No. of posts	\\ \hline
	Child unsafe	& 387		& 10,891	\\
	Untrustworthy	& 317		& 8,057		\\
	Questionable	& 312		& 8,859		\\
	Negative	& 266		& 5,863		\\
	Adult content	& 162		& 3,290		\\
	Spam		& 124		& 4,985		\\
	Phishing	& 39		& 495		\\ \hline
	Total		& 627		& 20,999	\\ \hline
	\end{tabular}
\caption{Number of malicious posts and pages in each category in our dataset. Number of pages posting phishing and spam URLs was the lowest.}
\label{tab:pagesPostsCatWise}
\end{center}
\end{table}

\section{Malicious pages on Facebook} \label{sec:analysis}

We now present our analysis and explore the characteristics of malicious and benign pages in detail. In particular, we identified and studied the most prominent entities posting malicious content, domain distribution of malicious domains, etc. We also studied and compared malicious and benign pages on the basis of their temporal behavior and network structure. %As mentioned in Section~\ref{sec:dataset}, all analysis in this section was done using the most recent 100 posts from all pages in our dataset.

\subsection{Entities posting malicious content}

\begin{figure}[!h]
     \begin{center}
                \subfigure[Word cloud for malicious pages. We found dominant presence of some politically polarized entities.]{%
                \label{fig:mal_tagcloud}
                \includegraphics[scale=0.30]{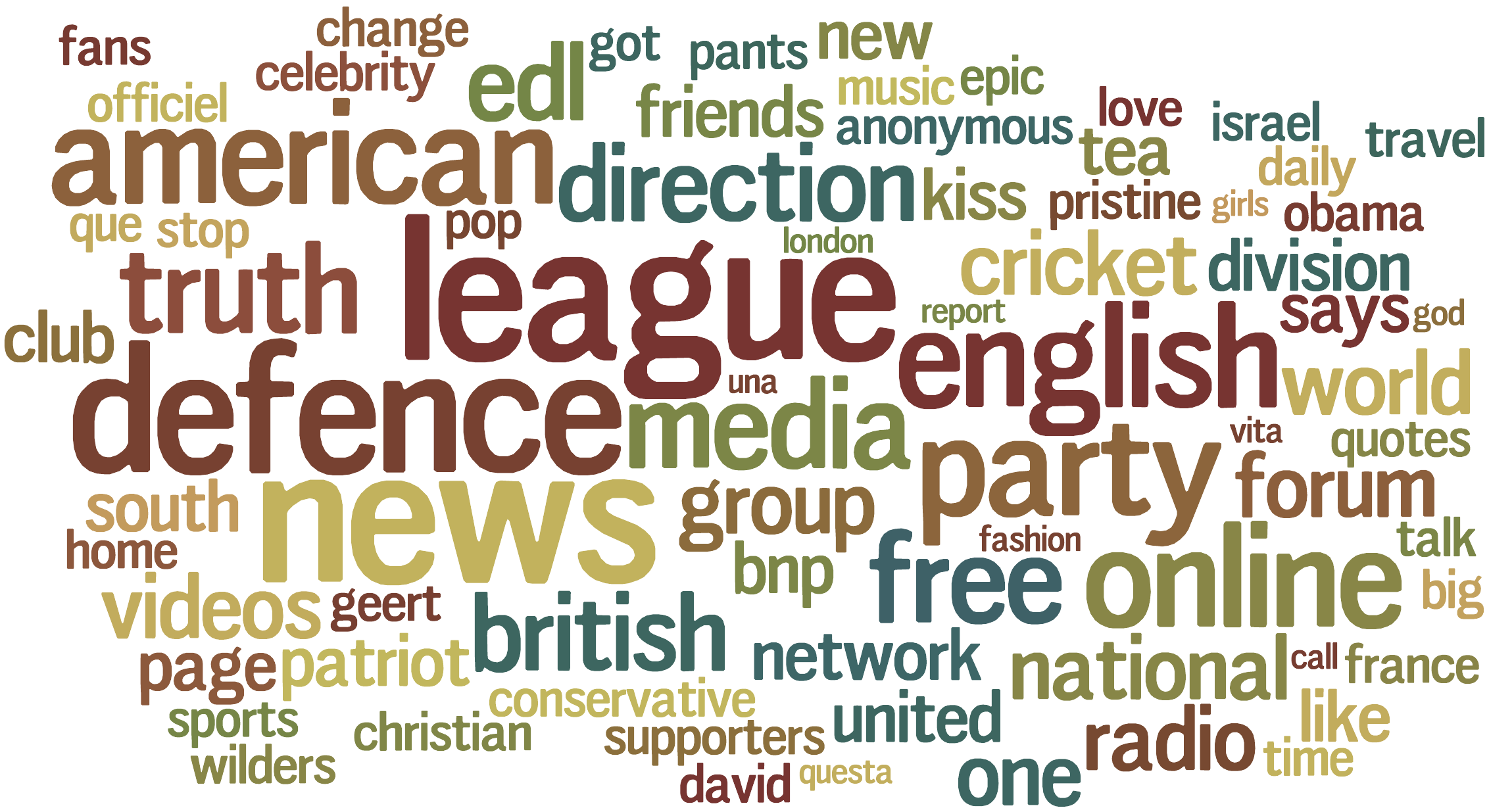}
        }
                \subfigure[Word cloud for benign pages. Words in benign page names were more generic.]{
                \label{fig:be9_tagcloud}
                \includegraphics[scale=0.30]{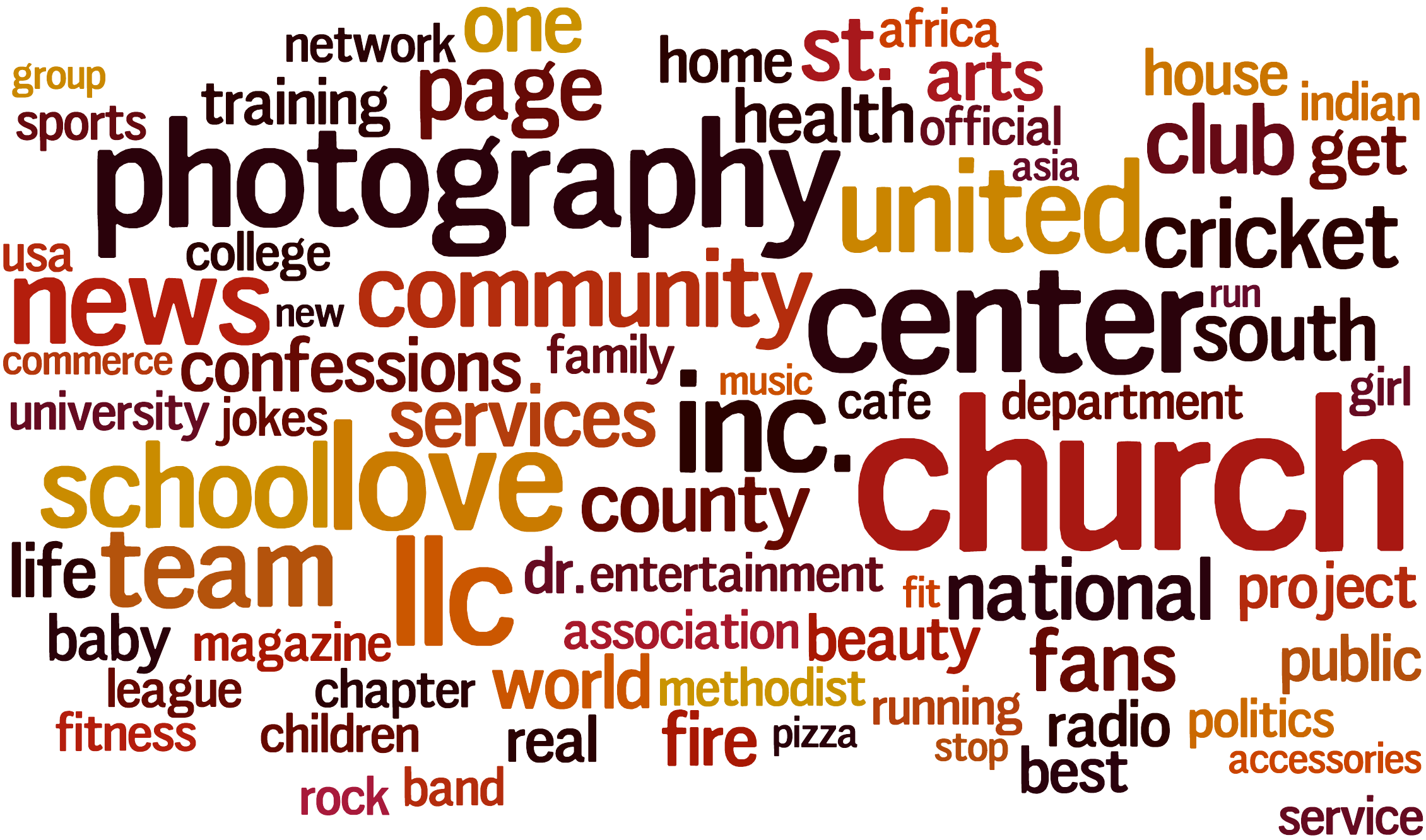}
        }
    \end{center}
\caption{Word clouds of the top 75 terms appearing in page names in our dataset. The size of the term is proportional to its frequency.}
   \label{fig:tagclouds}
\end{figure}

We first performed term-frequency analysis on unigrams, bigrams, and trigrams obtained from page names in our dataset to identify the most prominent entities generating malicious (and benign) content. Figure~\ref{fig:tagclouds} represents the top 75 unigrams obtained from page names. We found dominant presence of politically polarized entities and religious groups from keywords like \emph{american, british, patriot, defense, conservative, supporters, christian, united, etc.} in malicious pages (Figure~\ref{fig:mal_tagcloud}). Bigram and trigram analysis confirmed wide presence of entities like \emph{British National Party (BNP)}, \emph{The Tea Party}, \emph{English Defense League}, \emph{American conservatives}, \emph{Geert Wilders supporters}, etc. We also found some malicious pages dedicated to pop bands (One Direction), radio channels (Kiss FM), pages using \emph{anonymous} in their names, etc. We manually inspected all the aforementioned pages and validated that the page names were aligned with the content published by these pages, and were not misleading.

Using the bigram and trigram analysis, we divided pages belonging to politically polarized entities into four broad groups based on page names to help us study them better. These groups were i) America (9 pages), containing pages with ``america'' in their page name, ii) British National Party (7 pages), containing pages mentioning British National Party or BNP in their page name, iii) Conservative (6 pages), containing pages with the term ``conservative'' in the page name, and iv) Defence League (11 pages), containing pages using the phrase ``defence league'' in the page name. We manually verified each page to ensure that they fit in the group they were assigned. However, to maintain anonymity, we do not reveal the exact page names. We performed linguistic analysis on the content published by these 4 categories of pages separately using LIWC2007~\cite{pennebaker2007development}. LIWC is a text analysis software to assess emotional, cognitive, and structural components of text samples using a psychometrically validated internal dictionary. It determines the rate at which certain cognitions and emotions (for example, personal concerns like religion, death, and positive or negative emotions) are present in the text. LIWC has been widely used in the past to study social media content related to politics~\cite{stieglitz2012political,tumasjan2010election,tumasjan2010predicting}.

\begin{figure}[!h]
\begin{center}
\fbox{\includegraphics[scale=0.35]{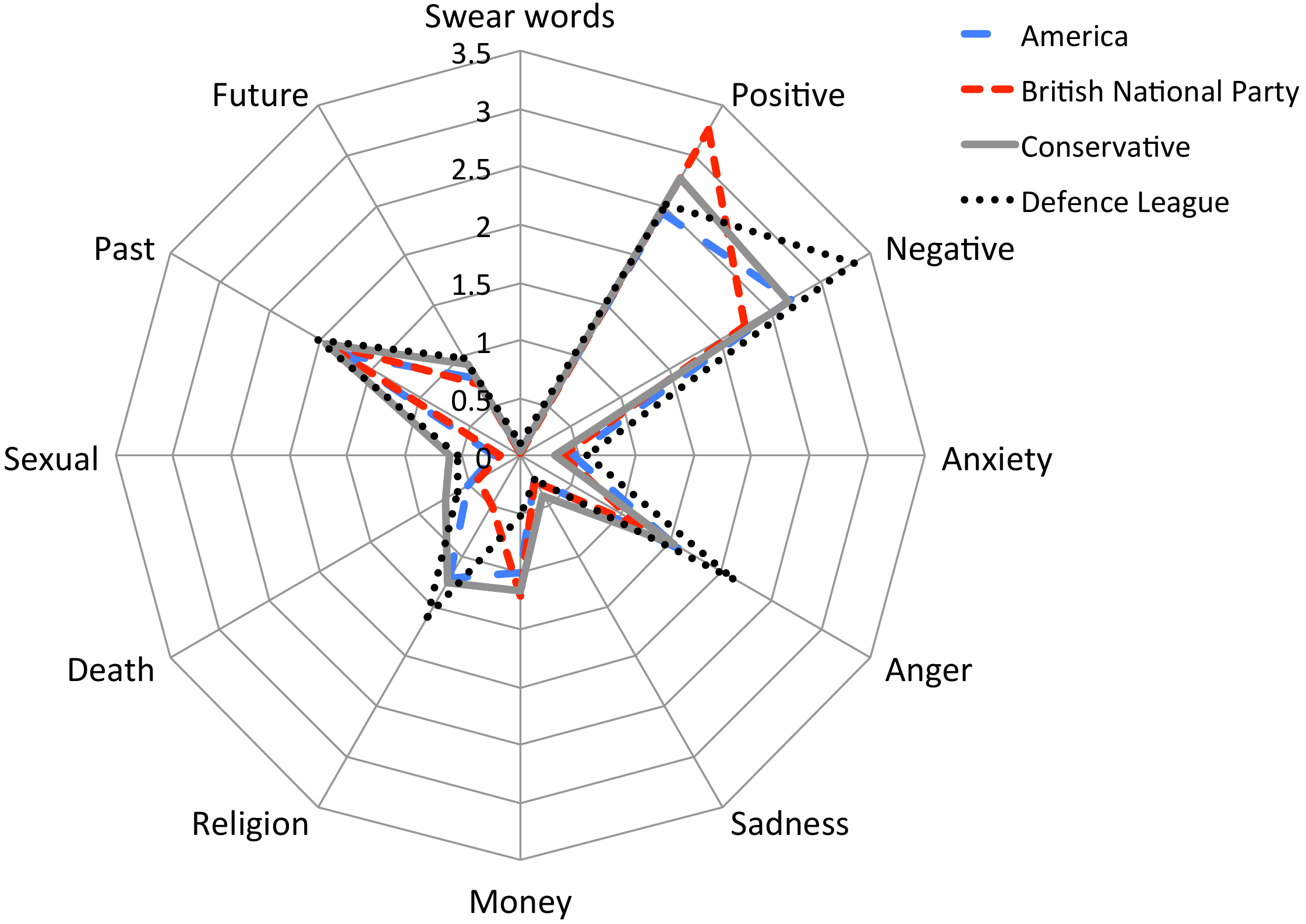}}
\end{center}
\caption{Linguistic analysis of content produced by politically polarized groups of pages in our dataset. We found high presence of negative emotion, anger, and religion related content.}
\label{fig:liwc_radar}
\end{figure}

We focused our analysis on 12 dimensions in order to profile the sentiment of content published by these groups of pages: Positive emotion, negative emotion, anxiety, anger, sadness, money, religion, death, sexuality, past orientation, future orientation, and swear words. Figure~\ref{fig:liwc_radar} shows the results of our analysis. We found high degree of anger in content from all categories. We also observed that only one category of pages (British National Party) had more positive emotions that negative emotions. The Defence League pages had much higher negative emotions as compared to positive emotions, followed by America pages. Conservative pages were almost equal in terms of positive and negative emotions. These findings contradicted prior results where researchers found that positive emotions outweighed negative emotions by 2 to 1 for profiles of all German political candidates~\cite{tumasjan2010predicting}. We also found substantial presence of content related to religion. These observations are indicative of the kind of influence that politically polarized pages in our dataset can have on their audience. To the best of our knowledge, no prior work in social network spam literature has looked at this class of malicious content.

Benign page names were found to represent a variety of categories and interests including \emph{photography, school, love, news, confessions, services}, etc., as shown in Figure~\ref{fig:be9_tagcloud} Bigram and trigram analysis revealed presence of a set of \emph{methodist church} pages. We also found some overlap between malicious and benign page names, for example, \emph{One Direction} fan pages, and radio channel pages. Unlike malicious pages, we did not find any fixed category of pages dominating in benign pages. 

\subsection{Domain distribution}
\begin{figure}[!h]
\begin{center}
\fbox{\includegraphics[scale=0.27]{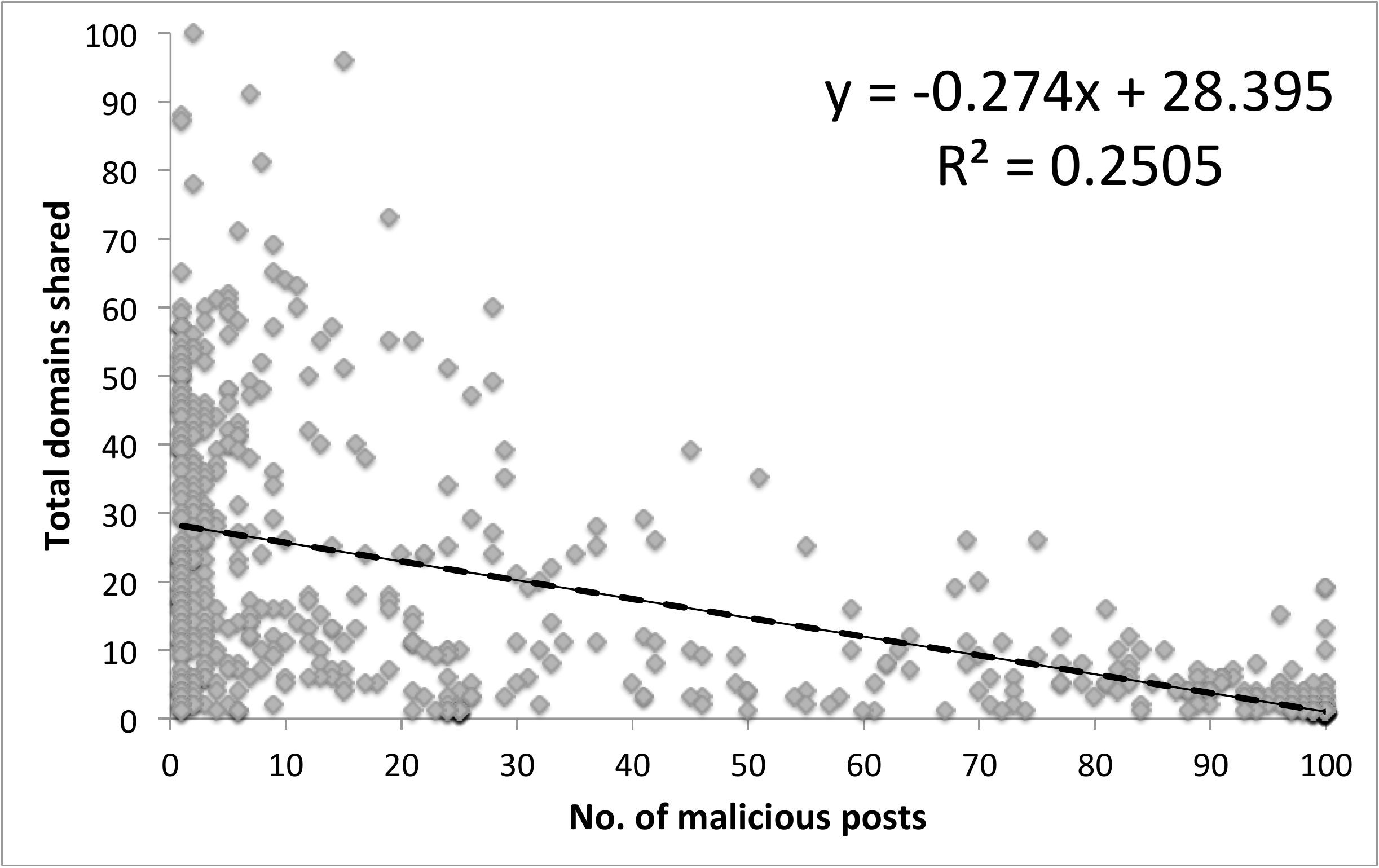}}
\end{center}
\caption{Number of malicious posts versus all domains published by all 627 pages in our dataset. We observed a weak declining trend in the total number of domains when the number of malicious posts published by a page increased.\protect\footnotemark}
\label{fig:postsVsDomains}
\end{figure}

Scanning the most recent 100 posts (as described in Section~\ref{sec:dataset}) revealed that almost half of the pages (49.28\%) in our dataset published 10 or less posts containing a malicious URL. Overall, the median number of domains shared by these pages was 24.5. On the contrary, the median number of domains shared by the other half of the pages posting more than 10 posts containing a malicious URL (50.72\%) was 5. Figure~\ref{fig:postsVsDomains} shows the distribution of the number of malicious posts versus the total number of domains shared by all malicious pages in our dataset. We found a weak declining trend in the number of domains as the number of malicious posts increased (r = -0.223, \emph{p-value}<0.01). 

\footnotetext{Three outliers (sharing 1,257, 202 and 140 domains) have been removed in this graph.}

This declining trend (and negative correlation) indicated that pages posting a large number of malicious URLs tend to do so from a small subset of domains. Infact, 84 pages in our dataset (13.39\%) shared URLs from only 1 domain. Out of these, 53 pages (8.45\%) published more than 90 posts containing a malicious URL, gathering \emph{likes} and \emph{comments} from 55,171 and 31,390 distinct users respectively. Most certainly, such pages exist for the sole purpose of promoting a single (malicious) domain, and are successful in engaging thousands of Facebook users. This sort of activity closely resembles social spam campaigns, which have been studied by multiple researchers in the past~\cite{gao2010detecting,grier2010spam,zhang2012detecting}. However, since most past research has focused on more obvious threats like unsolicited and targeted spam, advertising, and bulk messaging, other types of malicious content concerned with trustworthiness and child safety has largely remained unaddressed.

Note that there also exist multiple legitimate pages on Facebook dedicated to promote a single domain, for example, FIFA World Cup page (exclusively posting fifa.com URLs), BBC News page (exclusively posting bbc.com URLs), etc. We found 118 benign pages in our dataset (9.23\%) which were dedicated to promote a particular domain. Such behavior cannot therefore be associated exclusively with malicious activity. Malicious pages seem to take advantage of this fact and continue their activity, hiding in plain sight.

\paragraph{Top domains}
Table~\ref{tab:topdomains} lists the 10 most frequently occurring domains in our dataset of malicious pages, along with their WOT classification, Facebook audience, and Alexa world ranking.~\footnote{\url{http://www.alexa.com/}} For each domain, we calculated the number of posts the domain appeared in, the sum of \emph{likes}, \emph{comments}, and \emph{shares} on all these posts, the number of pages the domain appeared in, and the sum of \emph{likes} on all these pages. It was interesting to observe that 3 out of the top 10 domains were very famous, and were ranked within the top 3,000 domains worldwide on the Alexa ranking. Two of these domains were reported for being unsafe for children and spreading adult content. Although the Internet does not restrict the creation and promotion of adult and child unsafe content, most OSNs including Facebook have established community standards which restrict the display of adult and explicit content~\cite{Facebook.com:2015}. All of the other domains had low Alexa ranking worldwide. Only 3 of the top 10 domains were marked as spam, and none of the domains in the top 10 were reported for phishing or malware, signifying that untrustworthy and child unsafe content is much more prominent on Facebook than traditional forms of malicious content like spam and phishing.

\begin{table*}
\small
\begin{center}
    \begin{tabular}{p{2.1cm}|l|l|p{3.3cm}|l|p{1.1cm}|p{1cm}}
    \hline
    Domain            & WOT class, categories                          & Posts & Likes | comments | shares 	& Pages & Page likes & Alexa rank \\ \hline
    ammboi.com        & Untrustworthy, suspicious, spam, privacy risks & 456   & 666 | 61 | 195         	& 5     & 109,012    & 352,191    \\
    ridichegratis.com & Untrustworthy                                  & 424   & 428 | 14 | 252         	& 21    & 2,650,802  & -          \\
    blesk.cz          & Child unsafe, adult  content                   & 402   & 3,674 | 2,103 | 1,494       	& 8     & 864,554    & 2,924      \\
    says.com          & Child unsafe                                   & 386   & 387 | 15 | 62			& 5     & 97,784     & 27,684     \\
    ghanafilla.net    & Untrustworthy, scam, spam, suspicious          & 296   & 192 | 8 | 6			& 3     & 54,246     & 1,360,634  \\
    9cric.com         & Child unsafe                                   & 281   & 1,189 | 121 | 177		& 13    & 193,348    & 923,243    \\
    perezhilton.com   & Child unsafe, adult content                    & 274   & 26,088 | 3,516 | 1,128		& 8     & 1,701,834  & 2,192      \\
    nairaland.com     & Untrustworthy, misleading claims or unethical  & 201   & 238 | 89 | 31			& 3     & 116,131    & 1,329      \\
    pulsd.com         & Untrustworthy, child unsafe                    & 199   & 2 | 0 | 0			& 2     & 19,020     & 247,480    \\
    970wfla.com       & Spam                                           & 194   & 700 | 448 | 280		& 2     & 22,486     & 277,467    \\ \hline
    \end{tabular}
\caption{Top 10 malicious domains in our dataset with their Web of Trust classification, Facebook audience, and Alexa world rank. }
\label{tab:topdomains}
\end{center}
\end{table*}

The number of posts and pages associated with each of the top 10 domains revealed that there existed multiple Facebook pages dedicated to promoting all of these domains. We observed that all of the top 10 domains appeared in 2 or more pages, and two of the domains appeared in over 10 pages (ridichegratis.com in 21 pages; 9cric.com in 13 pages). At least 4 of the top 10 domains (ammboi.com, ghanafilla.net, pulsd.com, and 970wfla.com) had two or more Facebook pages each (3 for ghanafilla.net, 5 for ammboi.com), heavily promoting their respective domains (over 90 out of the 100 posts containing the domain, per page). Pages set up for these domains also had a substantial audience, with 6 out of the 10 domains collectively having over 100,000 \emph{likes} on their pages. Two of the top 10 domains had over 1 million \emph{likes} (collectively) on pages promoting them. The collective number of \emph{likes}, \emph{comments}, and \emph{shares} on posts was however, considerably low as compared to collective \emph{likes} on the pages. Only 3 out of the top 10 domains managed 1,000 or more \emph{likes} on the posts associated with them. This indicated that while malicious domains in our dataset were successful in gathering a substantial audience in the form of page \emph{likes}, they were not as successful in engaging the audience with their content. We also observed that 2 of the 3 domains with high Alexa rank (blesk.cz and perezhilton.com) also had high number of page \emph{likes} and high number of \emph{likes}, \emph{comments}, and \emph{shares} on posts. This signified that domains which were popular (more visited) on the Internet were also more famous on Facebook.

\subsection{Temporal behavior}

We explored the temporal activity of all pages in our dataset to determine how active the pages were. To be able to quantitatively compare the activity of malicious and benign pages, we calculated a \emph{daily activity ratio} for each page, defined by the ratio of number of days a page was active (published one or more posts) versus the total number of days between the first and hundredth post by the page.\\

$\emph{daily activity ratio} = \frac{\text{\emph{no. of days active}}}{\text{\emph{no. of days between first and last post}}}$ \\

\begin{figure*}[!ht]
     \begin{center}
		\subfigure[Hourly activity ratio of malicious and benign pages in our dataset.]{%
		\label{fig:fracHoursActive}
		\includegraphics[scale=0.3]{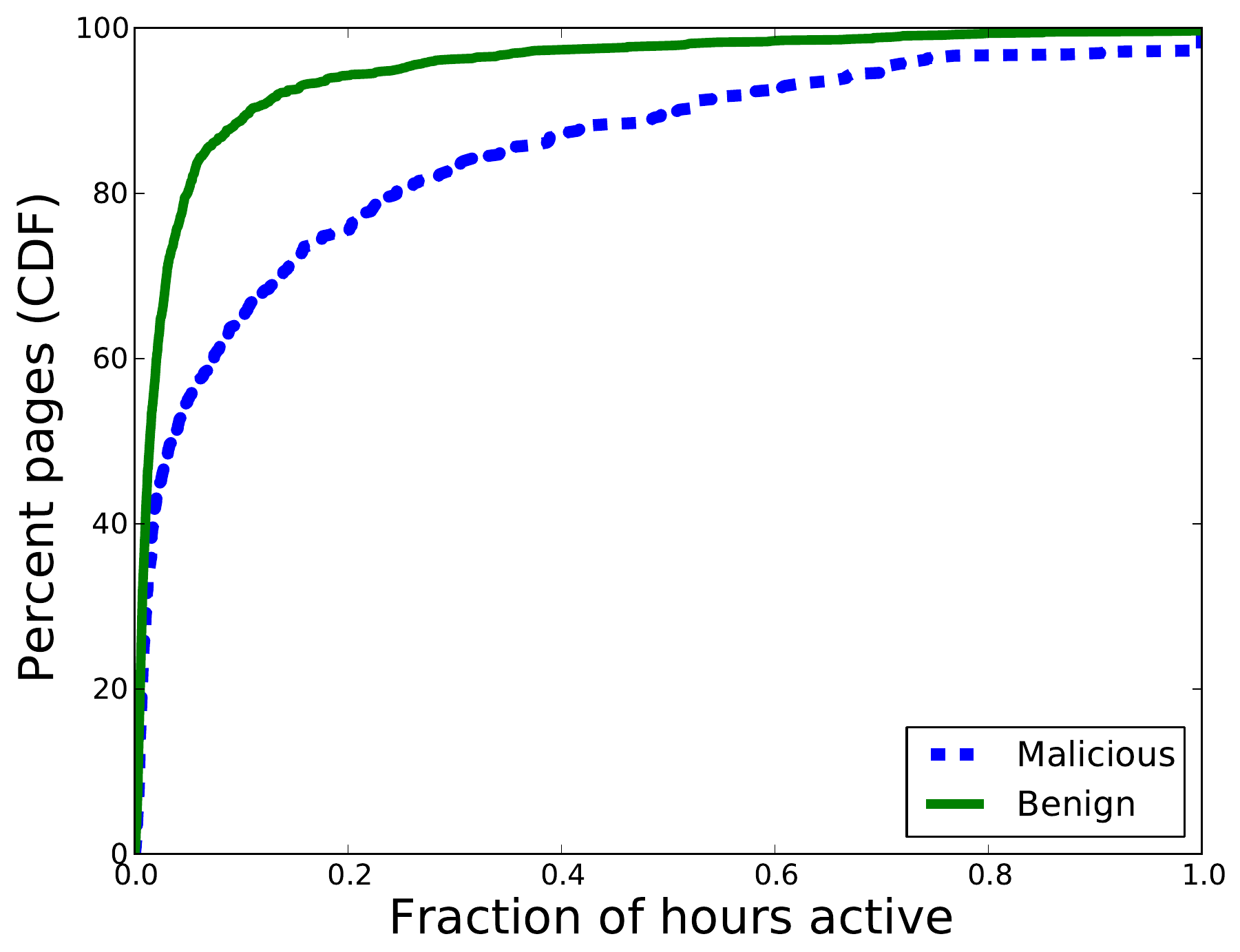}
        }\hspace{5pt}
		\subfigure[Daily activity ratio of malicious and benign pages in our dataset.]{%
                \label{fig:fracDaysActive}
                \includegraphics[scale=0.3]{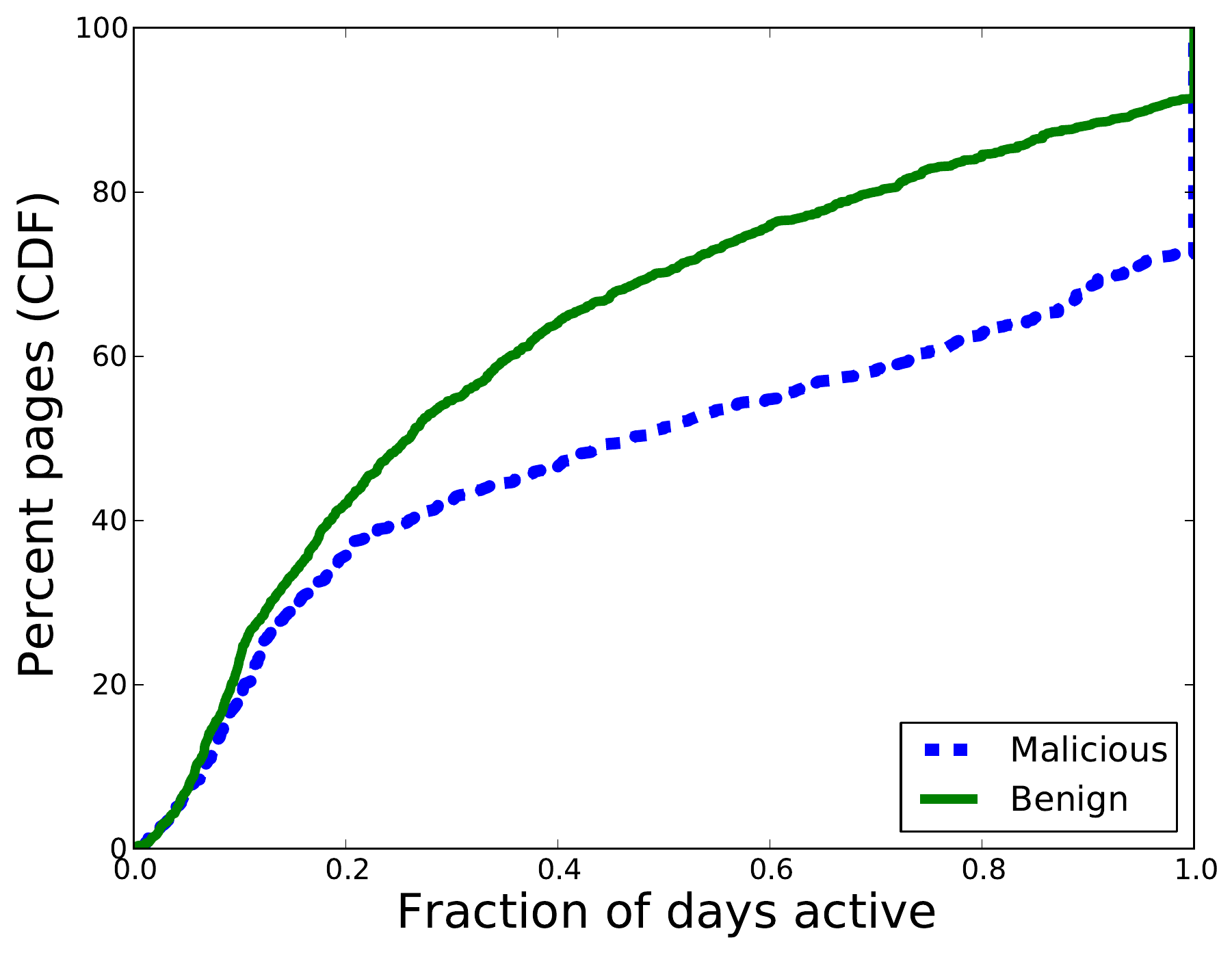}
        }\hspace{5pt}
		\subfigure[Weekly activity ratio of malicious and benign pages in our dataset.]{
                \label{fig:fracWeeksActive}
                \includegraphics[scale=0.3]{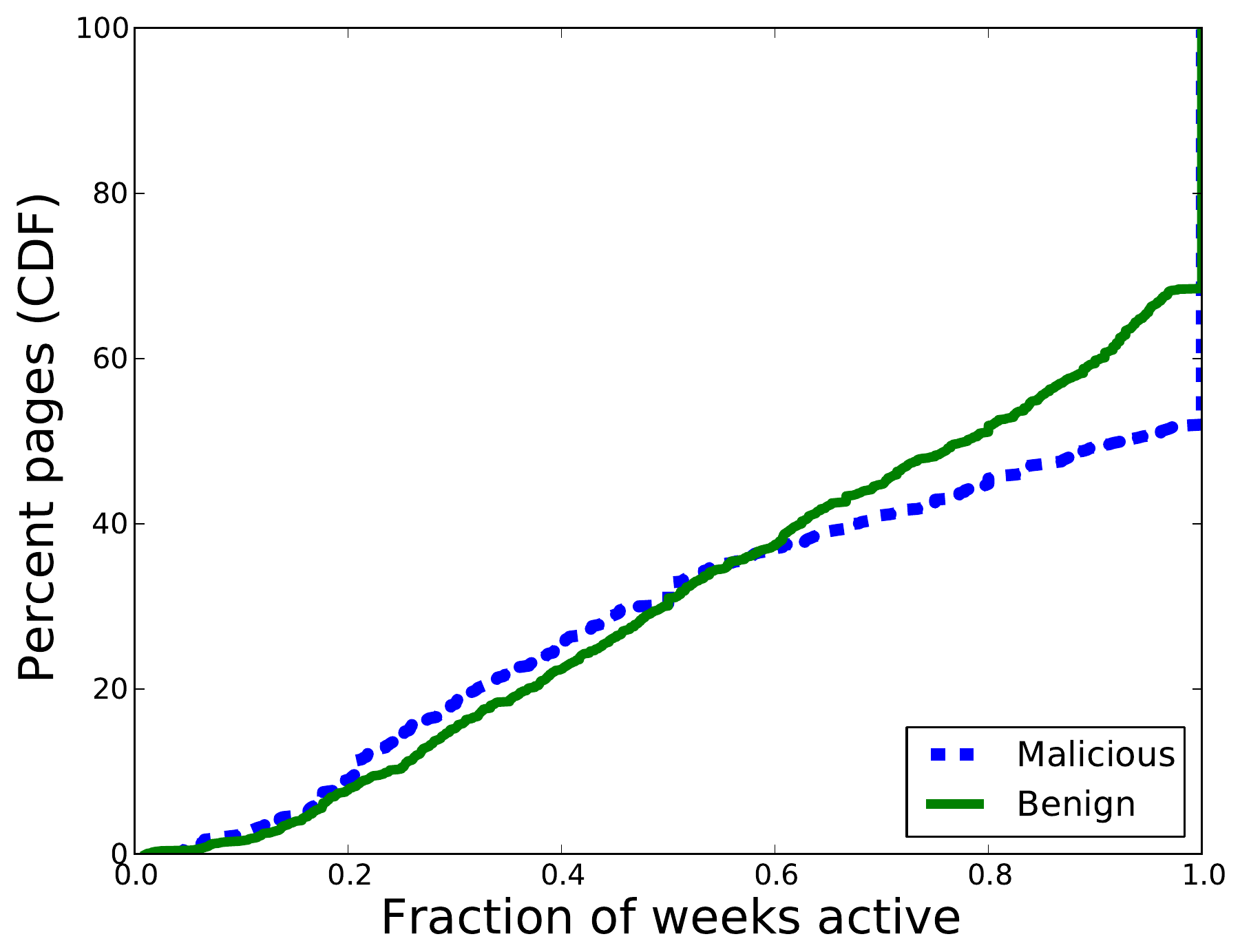}
	}
    \end{center}
\caption{Comparative analysis of daily, hourly, and weekly temporal activity of pages in our dataset. We found that malicious pages were more active than benign pages. }
   \label{fig:temporal}
\end{figure*}

Figure~\ref{fig:fracDaysActive} shows the \emph{daily activity ratio} plots of all malicious and benign pages in our dataset. We observed that 27.43\% of all malicious pages were active daily as compared to only 8.60\% daily active benign pages. On average, malicious pages were 1.4 times more active daily as compared to benign pages in our dataset. We also calculated activity ratio in terms on number of hours and number of weeks, and observed similar results. Table~\ref{tab:activityratio} shows the average hourly, daily, and weekly activity ratios of all malicious and benign pages in our dataset. Malicious pages were found to be 3 times as active as benign pages in terms of hourly activity (Figure~\ref{fig:fracHoursActive}), and 1.04 times as active in terms of weekly activity (Figure~\ref{fig:fracWeeksActive}). All activity ratio values were compared using Mann-Whitney U test and the differences were found to be statistically significant (\emph{p-value}<0.01 for all experiments)~\cite{mann1947test}. These difference confirmed that malicious pages in our dataset were more active as compared to benign pages, and published more frequently. 

\begin{table}[!h]
\begin{center}
    \begin{tabular}{l|l|l}
    \hline
    Activity ratio & Malicious & Benign \\ \hline
    Hourly         & 0.015     & 0.005  \\
    Daily          & 0.52      & 0.37   \\
    Weekly         & 0.73      & 0.70   \\ \hline
    \end{tabular}
\caption{Average activity ratios of malicious and benign pages in our dataset. Malicious pages were found to be more active than benign pages. }
\label{tab:activityratio}
\end{center}
\end{table}

\begin{figure}[!h]
     \begin{center}
                \subfigure[Time of the last post published by malicious and benign pages.]{
                \label{fig:lastpost}
                \includegraphics[scale=0.2]{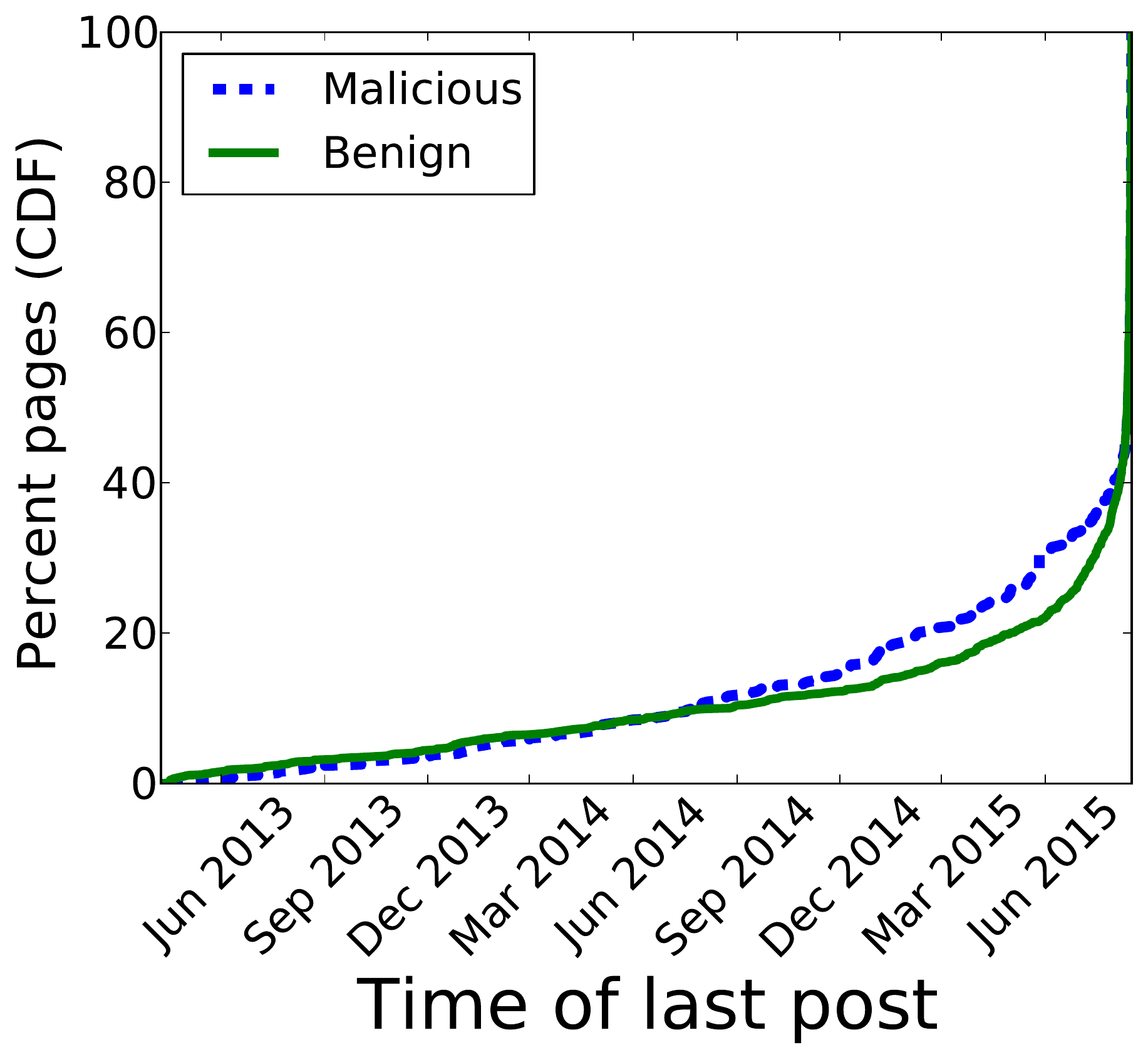}
        }
                \subfigure[Percentage of \emph{likes} gained by malicious pages over 2 months.]{%
                \label{fig:pagelikesovertime}
                \includegraphics[scale=0.2]{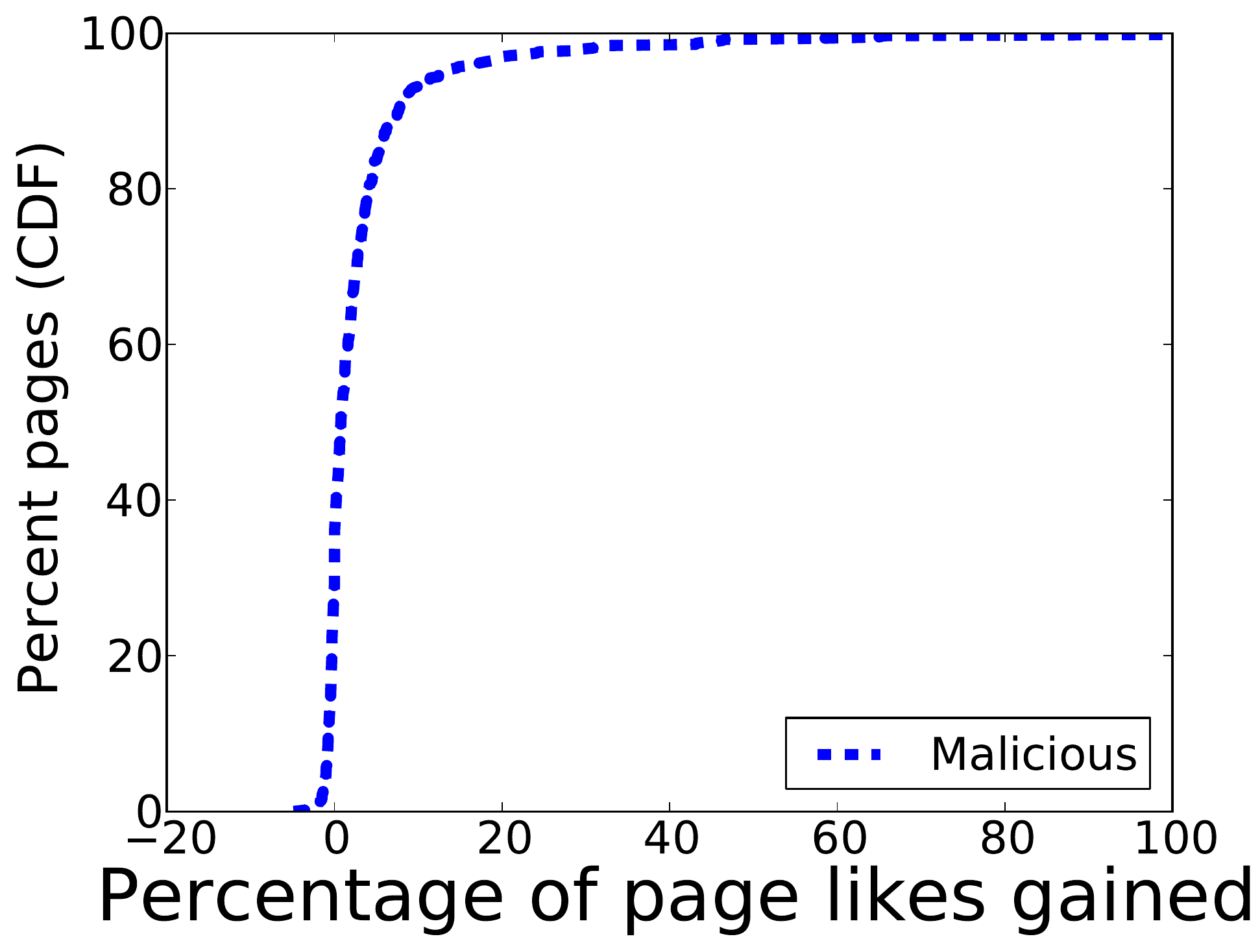}
        }
    \end{center}
\caption{Time of last post of malicious and benign pages, and \emph{likes} gained by malicious pages over time.}
   \label{fig:overtime}
\end{figure}

While studying the timestamps of posts published by pages, we came across some dormant pages that had not published a post in a long time. In literature, this behavior has been associated with malicious entities who set up OSN accounts to publish spam and propaganda just after a major event has taken place, and become inactive (or are suspended for suspicious behavior) soon after~\cite{gupta20131}. Figure~\ref{fig:lastpost} shows the timestamps of the last post generated by malicious and benign pages in our dataset. We calculated the average dormancy period (number of days between last post and data collection day), and did not find a substantial difference between malicious and benign pages ($\mu_{malicious}$ = 102.84 days, $\mu_{benign}$ = 91.16 days, Mann Whitney U test statistic = 381769.5, \emph{p-value}<0.05). We observed that 11.16\% malicious pages (and 10.01\% benign pages) in our dataset had been dormant for one year or more, while 20.41\% malicious pages (and 15.33\% benign pages) had not published a post in six months or more. These similar numbers for dormancy period of malicious and benign pages indicated the absence of any large group of malicious pages that may have been set up as part of a campaign, and became dormant together. However, there may exist some malicious pages in our dataset that were created right after an event to publish malicious content, and became inactive soon after.

\paragraph{Popularity over time}
We monitored the 627 malicious pages for two months, taking a snapshot of their page information every day between August 13 and October 13, 2015. Figure~\ref{fig:pagelikesovertime} shows the percentage of \emph{likes} gained by the pages over the two-month period. Over 90\% of the pages underwent a gain / loss of under 10\% in page \emph{likes}. We found a positive correlation (\emph{r}>0.7, \emph{p-value}<0.05) between page \emph{likes} and time for 343 pages (54.7\%). We also found negative correlation (\emph{r}<-0.7, \emph{p-value}<0.05) for 147 pages (23.1\%). The maximum gain for pages with increasing \emph{likes} was 99.95\%, while the maximum loss for pages with decreasing \emph{likes} over time was only 4.87\%. Likes on the rest of the pages were not strongly correlated with time. These numbers indicate that more (over double) number of malicious pages gained popularity over time as compared to the number of pages which lose popularity over time. Further, the gain far outweighs the loss. Visibly, malicious pages don't just exist, but gain popularity over time. This data collection is active. We intend to analyze the behavior of malicious pages and compare them with the behavior of benign pages over time in more detail in the future.

\subsection{Post and page metadata}

\begin{figure}[!h]
\begin{center}
\includegraphics[scale=0.5]{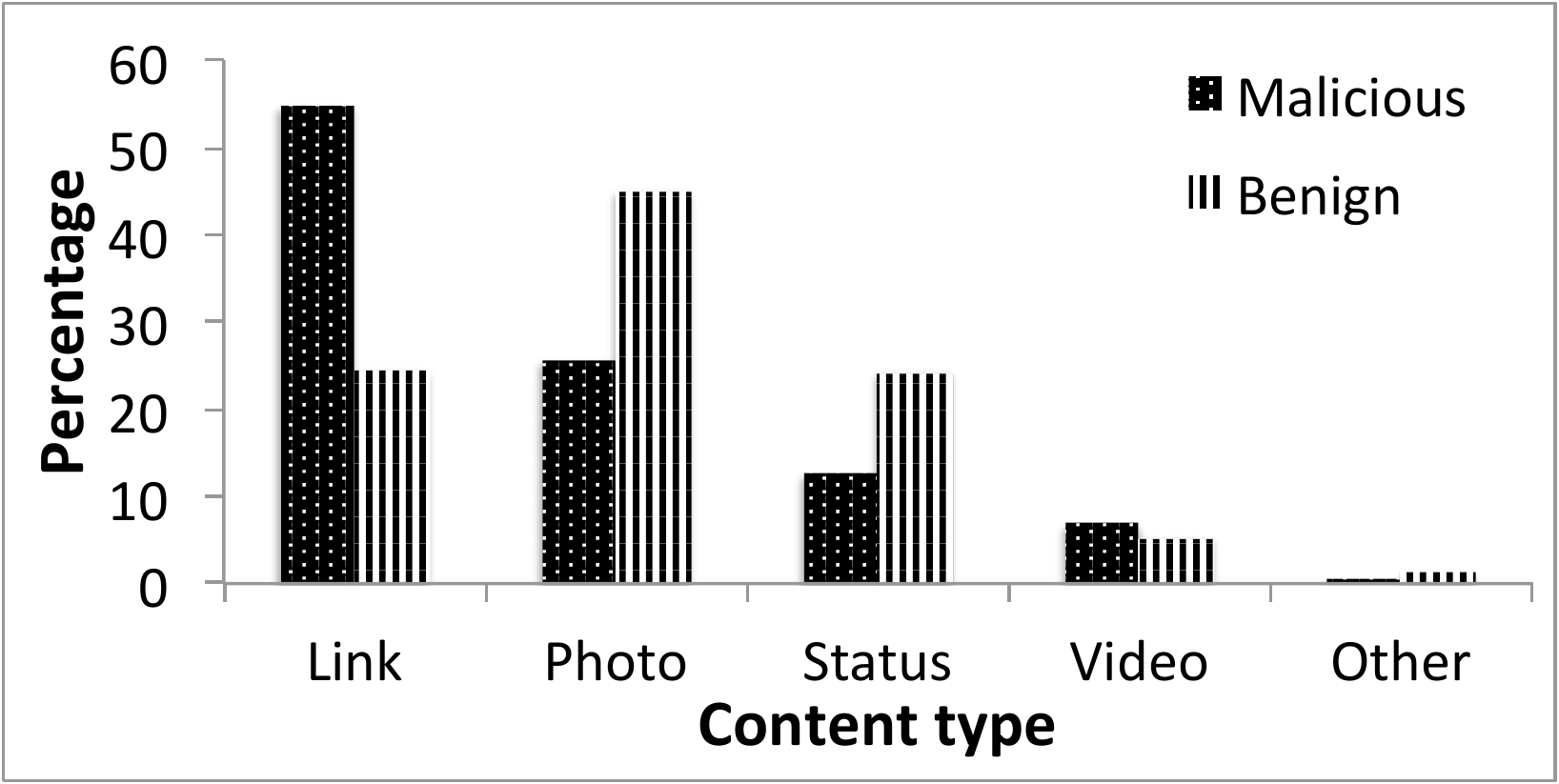}
\end{center}
\caption{Types of content published by malicious and benign pages in our dataset. Malicious pages published more links, while benign pages published more pictures and videos.}
\label{fig:postTypes}
\end{figure}

Analyzing the metadata of posts in our dataset revealed some significant differences in the type of content published by malicious and benign pages. Figure~\ref{fig:postTypes} shows the distribution of the content type of posts published by all pages in our dataset. We observed that more than half of the content published by benign pages were photos and videos (50.16\%). This percentage went down to 32.42\% for malicious pages. The metadata also revealed that over half of the posts published by malicious pages were links (54.69\%), where as less than a quarter of all posts published by benign pages were links (24.45\%). These numbers indicate that malicious pages are inclined towards posting links, and directing user traffic to external websites. On the other hand, benign pages tend to post more pictures and videos, which can be consumed by users without leaving the OSN. In addition to content types, we looked at the status types of posts and found that benign pages published almost double the amount of content through mobile devices (23.80\%) as compared to malicious pages (12.33\%). 

All pages on Facebook have a \emph{category} associated with them, for example \emph{Community, Company, Personal Website}, etc. This category is assigned to the page by the page administrator(s) at the time of page creation, according to the person / organization represented by the page, and content that the page generates. To see if any subset of categories was more popular among a particular class of pages (malicious or benign), we compared category ranks and found strong correlation between category ranks across malicious and benign pages (Spearman's $\rho$ = 0.67, \emph{p-value}<0.01). This indicated that the distribution of malicious and benign pages across various categories was fairly similar, and that categories more popular among malicious pages were also more popular among benign pages. We also compared the page \emph{likes} and page mentions (\emph{talking\_about\_count} field) of malicious and benign pages, and did not find any significant differences. 

These observations indicated that apart from the type and source of published content, there were no significant differences in the meta information between malicious and benign pages in our dataset. Metrics like popularity (\emph{likes}) and user mentions (\emph{talking\_about\_count}) associated with OSN entities can be used to identify spammers, since they capture the notion of influence of entities in the network~\cite{cha2010measuring}. However, similarities in such metrics across malicious and benign pages can aid malicious pages to continue operating regularly and go undetected for long periods of time, hiding in plain sight.

\subsection{Network analysis}

Past research has shown that decentralized networks are prone to \emph{sybil attacks}, wherein malicious entities tend to collude together and attempt to infiltrate the legitimate part of the network~\cite{douceur2002sybil}. Such attacks have also been studied in context of other OSNs~\cite{yang2014uncovering}. To investigate if such phenomenon exists for Facebook pages too, we analyzed the \emph{likes, comments}, and \emph{shares} networks for both malicious and benign pages in our dataset. Facebook does not provide an API endpoint to gather the list of users who have \emph{liked} (subscribed to) a page. However, it is possible to collect the list of users who have \emph{liked, commented} on, or \emph{shared} posts originating from a page. As described in Section~\ref{sec:dataset}, we collected all \emph{likes, comments}, and \emph{shares} on the most recent 100 posts of all pages in our dataset, and analyzed the inter and intra-page networks. In particular, we analyzed networks consisting of pages and users \emph{liking, commenting on,} or \emph{sharing} posts from two or more pages in our dataset (malicious and benign separately) (inter-page networks), and networks of pages \emph{liking, commenting on,} or \emph{sharing} posts from pages within our dataset (malicious and benign separately) (intra-page networks). To keep the network size comparable, we averaged out the results for 10 random samples of 627 benign pages each (same size as malicious pages dataset) drawn from the entire dataset of 1,278 benign pages. We used Gephi for all our analysis~\cite{bastian2009gephi}.

Table~\ref{tab:lcs_networks} shows the details of the network analysis. We found that the Inter-likes network for benign pages (83,799 nodes) was much larger and stronger (avg. weighted degree: 41.695) than the Inter-likes network for malicious pages (21,947 nodes, avg. weighted degree: 24.273), indicating that a larger number of users \emph{liked} two or more benign pages as compared to the number of users who \emph{liked} two or more malicious pages in our dataset. More interestingly, we found stronger ties (avg. weighted degree for Intra-page networks) within malicious pages in all aspects (\emph{likes, comments}, and \emph{shares}) as compared to benign pages, indicating collusion and sybil behavior within malicious pages. We also found a much larger number of communities in all Inter-page networks for benign pages as compared to Inter-page networks for malicious pages, indicating a larger and more diverse audience for benign pages as compared to malicious pages.

\begin{table*}[!ht]
    \begin{tabular}{l|p{1cm}|p{1cm}|p{2cm}|p{1cm}|p{1.5cm}|p{1.6cm}|p{2.6cm}}
    \hline
    Network type		& Total nodes	& Total edges	& Avg. weighted degree	& Density & Modularity & No. of communities	& Weakly connected components \\ \hline
    \multicolumn{8}{c}{Malicious (All 627 pages)} \\ \hline
    Inter-page \emph{likes} network    & 21,947	& 103,683	& 24.273		& 0       & 0.492	& 18			& 2	\\
    Inter-page \emph{comments} network & 3,901		& 13,957	& 11.255		& 0.001   & 0.607	& 19			& 2	\\
    Inter-page \emph{shares} network   & 14,318	& 67,513	& 15.796		& 0       & 0.480	& 14			& 1	\\ \hline
    Intra-page \emph{likes} network    & 27		& 35		& 8.333			& 0.05    & 0.389	& 9			& 9	\\
    Intra-page \emph{comments} network & 9		& 9		& 1.667			& 0.125   & 0.551	& 3			& 3	\\
    Intra-page \emph{shares} network   & 68		& 65		& 6.309			& 0.014   & 0.705	& 21			& 21	\\ \hline \hline
    \multicolumn{8}{c}{Benign (Results averaged across 10 random samples of 627 benign pages each)} 							\\ \hline
    Inter-page \emph{likes} network	& 83,799	& 390,854	& 41.695		& 0	  & 0.391	& 3070			& 1	\\
    Inter-page \emph{comments} network & 2,958		& 7,722		& 8.919			& 0.001	  & 0.595	& 142			& 2	\\
    Inter-page \emph{shares} network   & 3,406		& 10,234        & 9.920                 & 0.001   & 0.620	& 30			& 1	\\ \hline
    Intra-page \emph{likes} network    & 4.3		& 3.6		& 0.408			& 0.075   & 0.079	& 0.7			& 0.7	\\
    Intra-page \emph{comments} network & 0		& 0             & 0                     & 0       & 0		& 0			& 0	\\
    Intra-page \emph{shares} network   & 7.8		& 6.9		& 1.168			& 0.072   & 0.175	& 1.1			& 1.1	\\ \hline
    \end{tabular}
\caption{Network analysis of \emph{likes}, \emph{comments} and \emph{shares} networks within and between pages in our dataset. We observed that malicious pages had stronger intra-network ties as compared to benign pages.}
\label{tab:lcs_networks}
\end{table*}

\begin{figure*}[!ht]
     \begin{center}
                \subfigure[Malicious pages \emph{liking} each others' posts; intra-likes network.]{
                \label{fig:intraLikesNetwork}
                \fbox{\includegraphics[scale=0.25]{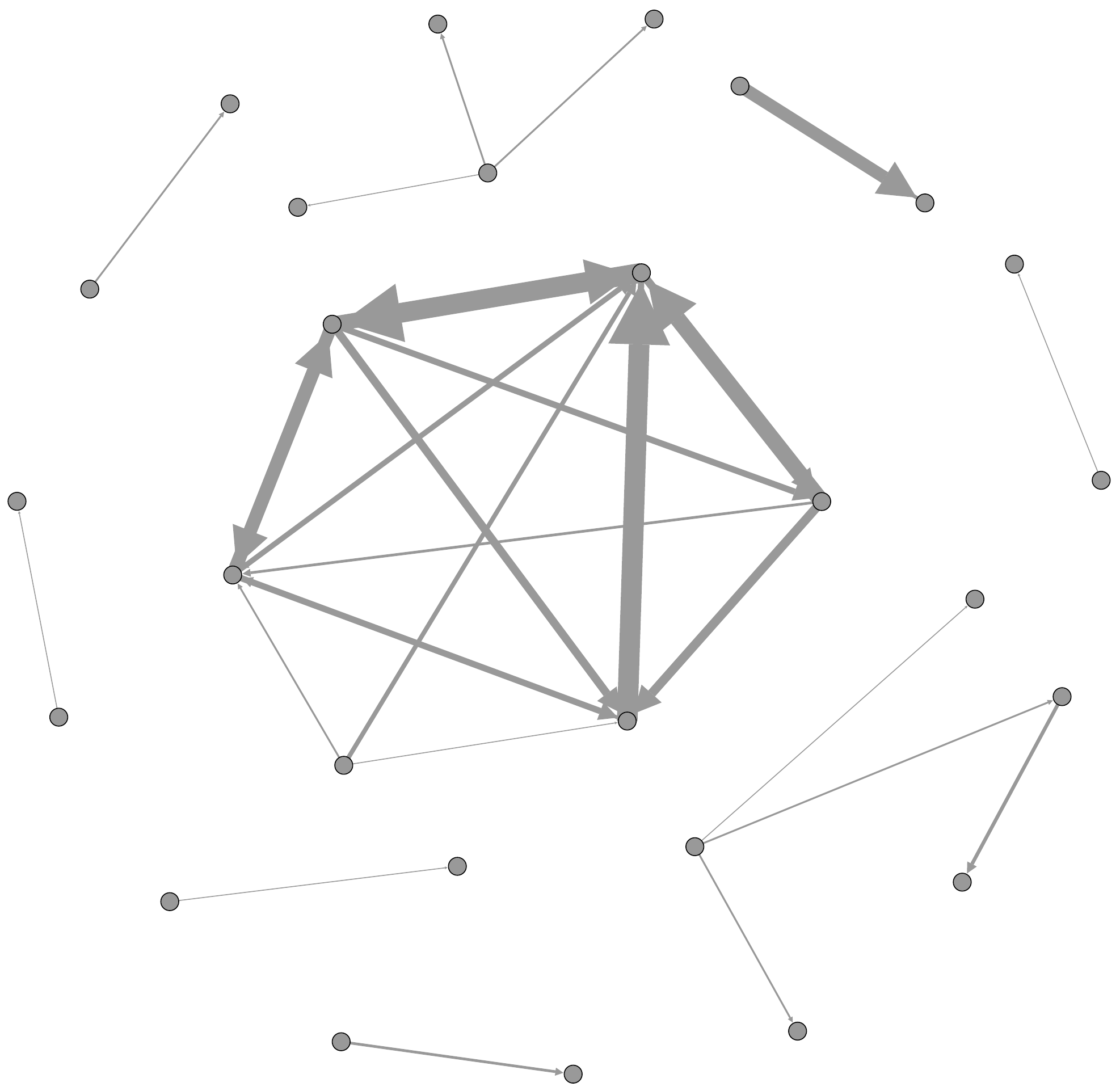}}
        }\hspace{10pt}
                \subfigure[Malicious pages \emph{commenting} on each others' posts; intra-comments network.]{
                \label{fig:intraCommentsNetwork}
                \fbox{\includegraphics[scale=0.25]{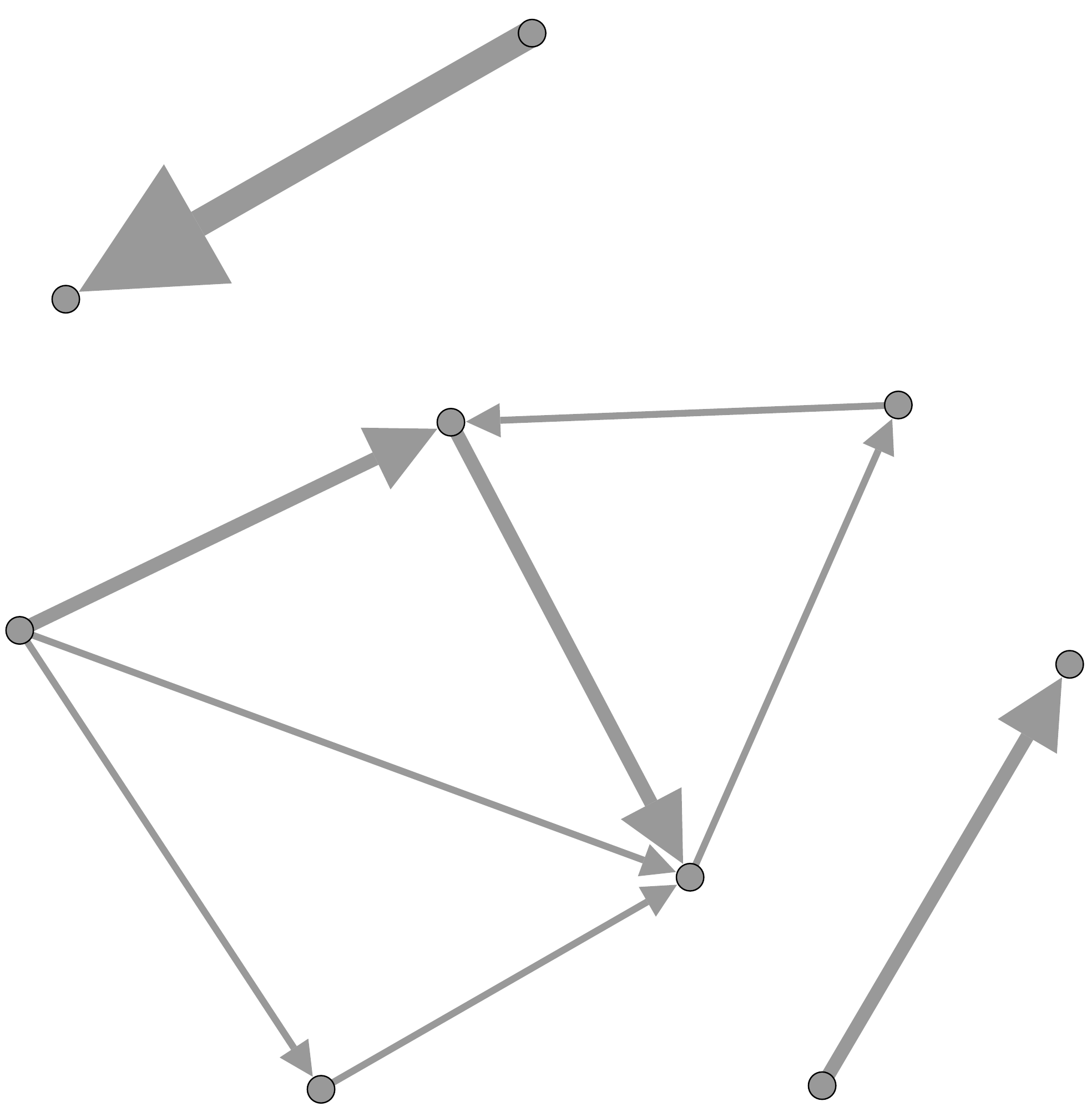}}
        }\hspace{10pt}
                \subfigure[Malicious pages \emph{sharing} each others' posts; intra-shares network.]{
                \label{fig:intraSharesNetwork}
                \fbox{\includegraphics[scale=0.25]{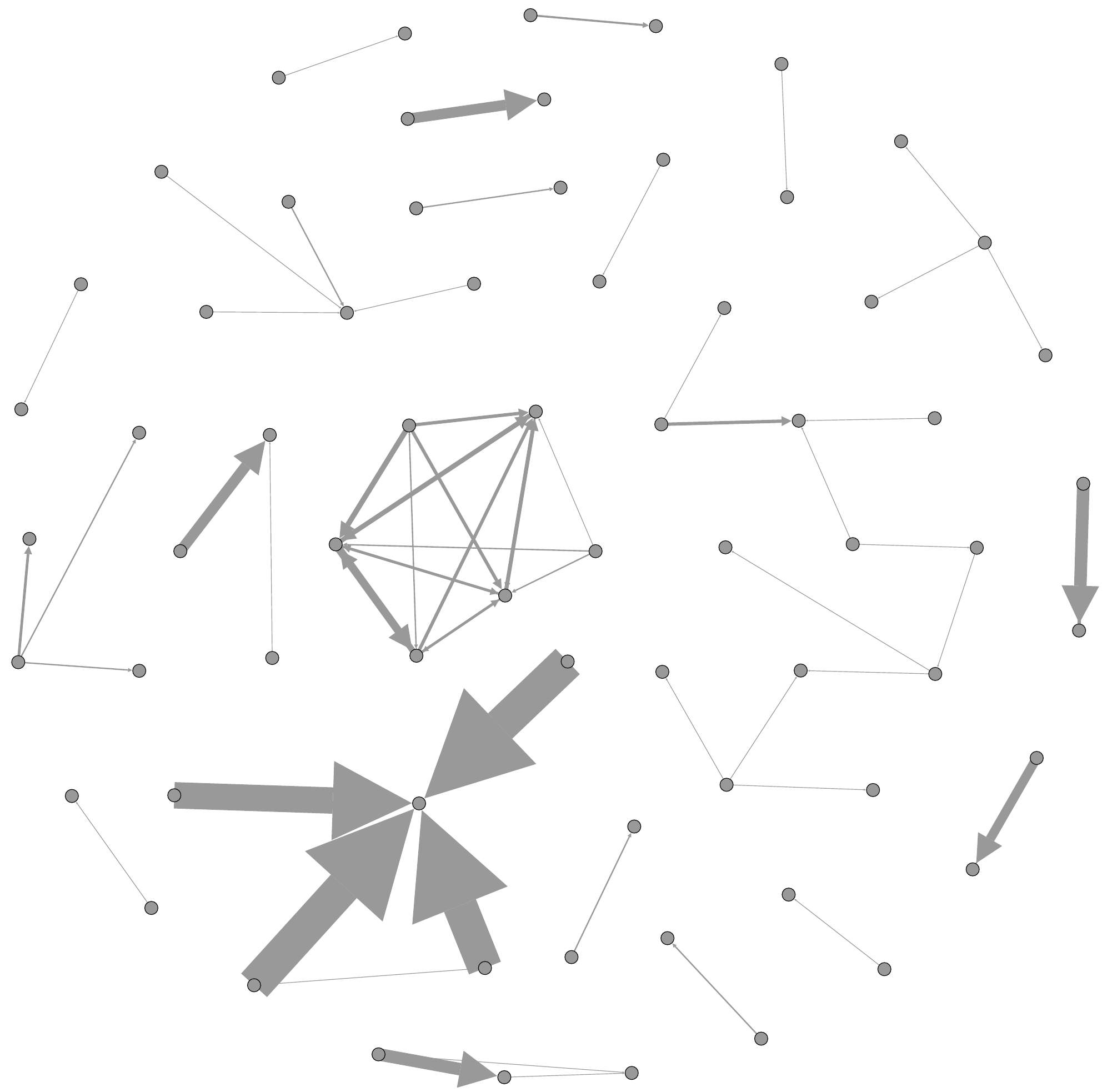}}
        }
    \end{center}
\caption{Network graphs capturing intra-page activity within malicious pages in our dataset. We found multiple two-node communities and a few bigger communities.}
   \label{fig:networks}
\end{figure*}

Stronger ties within malicious pages prompted us to further investigate the communities we detected from Intra-page \emph{likes, comments}, and \emph{shares} networks. Figure~\ref{fig:networks} shows the network graphs of the detected communities. We observed that post \emph{sharing} was the most prominent intra-page activity, followed by \emph{liking} and \emph{commenting}. The network graphs also revealed a distinct community of six Facebook pages completely connected to each other in terms of \emph{likes} (Figure~\ref{fig:intraLikesNetwork}) and \emph{shares} (Figure~\ref{fig:intraSharesNetwork}). Five out of these six pages also formed a community in the intra-comments graph (Fig~\ref{fig:intraCommentsNetwork}). All pages in this community belonged to adult stars and promoted pornographic content. This behavior closely resembled a sybil network, and strongly indicated that all these pages were controlled by / belong to the same real-world entity (person or organization). We also found multiple two-page communities involving politically polarized pages, where one page heavily engaged in \emph{liking, commenting on}, and \emph{sharing} the other page's content. We intentionally do not provide any information about these pages (page name, ID, etc.) to maintain anonymity.

%\subsection{Verified pages}

\section{Discussion} \label{sec:discussion}

In this section, we discuss the implications of our analysis and results. We also discuss potential directions for new techniques to curb the spread of malicious content on OSNs.\\

{\bf Politically polarized entities:} Our analysis revealed the presence of some politically polarized entities in our dataset of malicious pages. We do not intend to implicate such entities from our findings. Entities involved in politics tend to be followed by masses with similar orientation, and is a global phenomenon in the real world. It is likely that such activity exists on online platforms other than Facebook too. We do not propose to debar such activity. However, we believe that extremely polar content should be moderated both online and offline, in order to maintain stability among the masses. An easy way to moderate such entities can be to display nudges or warning messages to users before they subscribe to such pages on any online platform~\cite{wang2013privacy}.

{\bf Beyond pages:} Pages on Facebook have a lot in common with Facebook groups and events. Groups and events can also be used to target large audiences at once. Moreover, Facebook has a common definition of ``Page Spam'' for pages, groups and events, and explicitly states that \emph{Pages, groups or events that confuse, mislead, surprise or defraud people on Facebook are considered abusive.} Our analysis and results can thus be easily extended to study malicious groups and events as well.

{\bf Automatic identification of malicious pages:} Our findings shed some light on various differences (like temporal behavior, content type, etc.) between malicious and benign pages, which can be used to train machine learning models to automatically differentiate between malicious and benign pages. These findings, however, are based on a limited history (up to 100 posts) of page activity. Although it is possible to collect and analyze the entire history for all pages, doing so would be time consuming and computationally expensive. Analyzing complete page history will also be prone to missing out on smaller, subtle behavioral patterns while looking at the bigger scheme of things. Moreover, pages can change behavior over time; malicious pages may stop spreading malicious content, while benign pages may start engaging in posting malicious content over time. In order to be effective in the modern era, machine learning based solutions need to be quick, real time and robust. To accommodate such changes in behavior, we recommend a self-adaptive model which relies partly on page information and partly on the recent activity by the page. The amount of history (number of posts) to consider can be decided experimentally. Such a model would be accommodative of the changing behavior of pages over time, and may remain accurate for long without the need for re-training.

\section{Related Work} \label{sec:relatedwork}

Multiple researchers have studied and proposed techniques to detect malicious content on Facebook and other OSNs in the past. Gao et al. presented an initial study to quantify and characterize spam campaigns launched using accounts on Facebook~\cite{gao2010detecting}. They studied a large anonymized dataset of 187 million asynchronous ``wall" messages between Facebook users, and used a set of automated techniques to detect and characterize coordinated spam campaigns. Authors of this work relied on URL blacklists to detect spam URLs and concentrated on spam, phishing and malware. Following up their work, Gao et al. presented an online spam filtering system that could be deployed as a component of the OSN platform to inspect messages generated by users in real time~\cite{gao2012towards}. In an attempt to protect Facebook users from malicious posts, Rahman et al. designed a social malware detection method which took advantage of the social context of posts~\cite{rahman2012efficient}. Authors were able to achieve a maximum true positive rate of 97\%, using a SVM based classifier trained on 6 features. This model was then used to develop MyPageKeeper~\footnote{\url{https://apps.facebook.com/mypagekeeper/}}, a Facebook app to protect users from malicious posts. Similar to Gao et al's work~\cite{gao2010detecting}, this work was also targeted at detecting spam campaigns.

Stringhini et al.~\cite{stringhini2010detecting} utilized a honeypot model to collect information about spammers on Facebook. They created and monitored a honey profile for over one year, and manually identified 173 spam profiles among a total of 3,831 friendship requests they received. This technique, however, is not extendable to Facebook pages. Ahmed et al. presented a Markov Clustering (MCL) based approach for the detection of spam profiles on Facebook. Authors crawled the public content posted by 320 handpicked Facebook users, out of which, 165 were manually identified as spammers, and 155 as legitimate. Authors then extracted 3 features from these profiles, viz. Active friends, Page Likes, and URLs to generate a weighted graph, which served as input to the Markov Clustering model~\cite{ahmed2012mcl}. Thomas et al. identified and studied over 1.1 million accounts suspended by Twitter for disruptive activities. Authors identified an emerging marketplace of illegitimate programs operated by spammers that included Twitter account sellers, ad-based URL shorteners, and spam affiliate programs that helped enable underground market diversification. Their results showed  that 77\% of spam accounts identified by Twitter were suspended within one day of their first tweet~\cite{thomas2011suspended}. Grier et al. also studied spam on the Twitter network and found that 8\% of 25 million URLs posted to the site pointed to phishing, malware, and scams listed on popular blacklists~\cite{grier2010spam}.

Most aforementioned research relied on URL blacklists to identify ground truth spam, phishing, and malware, and tried to identify patterns which could be used to design effective measures to curb the spread of spam on OSN platforms. However, fewer attempts have been made to go beyond the traditional spam, phishing, and malware, and address other classes of malicious content on OSNs which include untrustworthy content, hate and discrimination, etc. that are non-trivial to identify through automated means. There has been some research in the space of identifying credible content on Twitter~\cite{castillo2011information,gupta2014tweetcred}, but state-of-the-art techniques proposed by researchers to detect content credibility have not been able to achieve the degree of efficiency that has been achieved in detecting traditional spam, phishing and malware.

\section{Conclusion and Future work} \label{sec:conclusion}

OSNs are full of a wide range of hostile entities that promote and spread various types of malicious content. In this paper, we identified and characterized Facebook pages posting malicious URLs. We moved beyond traditional types of malicious content like unsolicited bulk messages, spam, phishing, malware, etc. that have been widely studied in the past, and studied a wider range of content that is deemed as malicious by community standards and Page Spam definitions established by Facebook. We focused our analysis on Facebook pages because of their public nature, vast audience, and inflated malicious activity~\cite{dewan2015towards}. Our observations revealed significant presence of politically polarized entities among malicious pages. Further, we found a substantial number of malicious pages dedicated to promote content from a single malicious domain. We also observed that malicious pages were more active than benign pages in terms of hourly, daily, and weekly activity. There existed some famous domains amongst the 10 most popular malicious domains in our dataset which were reported for child unsafe and adult content. Network analysis revealed possible presence of collusive behavior among malicious pages that engaged heavily in promoting each others' content. To the best of our knowledge, this is one of the first studies focused on characterizing Facebook pages posting malicious content. We believe that our findings will enable researchers to better understand the landscape of malicious Facebook pages that have been hiding in plain sight and promoting malicious content seemingly unperturbed.

In future, we would like to do a detailed longitudinal study on malicious Facebook pages to characterize their behavior over time. We also plan to use our findings to devise efficient automated techniques and develop technological solutions to enable end users to identify malicious Facebook pages. We would also like to expand our analysis to identify malicious ``groups'' and ``events'' on Facebook which largely remain unexplored to date. 

\bibliographystyle{abbrv}
\bibliography{/Users/prateekdewan/Dropbox/Super_BibTex_Collection/Prateek}

\end{document}